\documentclass[sigconf,nonacm]{acmart}
\usepackage{xspace}
\usepackage[svgnames]{xcolor}
\usepackage[inline,shortlabels]{enumitem}
\usepackage{algorithm}
\usepackage{algpseudocode}
\usepackage{multirow}
\usepackage{custom}
\AtBeginDocument{%
  }

\acmConference[]{}{}{}

\begin{document}

\title[VILLA]{VILLA: \underline{V}ersatile \underline{I}nformation Retrieval From Scientific \underline{L}iterature Using Large \underline{La}nguage Models}

\author{Blessy Antony}
\orcid{0000-0002-1269-5191}
\affiliation{%
  \institution{Virginia Tech}
  \city{Blacksburg}
  \state{Virginia}
  \country{USA}
}
\email{blessyantony@vt.edu}

\author{Amartya Dutta}
\author{Sneha Aggarwal}
\affiliation{%
  \institution{Virginia Tech}
  \city{Blacksburg}
  \state{Virginia}
  \country{USA}
}
\email{{amartya, sengg}@vt.edu}


\author{Vasu Gatne}
\affiliation{%
  \institution{Virginia Tech}
  \city{Blacksburg}
  \state{Virginia}
  \country{USA}
}
\email{gatne@vt.edu}

\author{Ozan Gökdemir}
\orcid{0000-0001-5299-1983}
\affiliation{%
  \institution{University of Chicago}
  \city{Chicago}
  \state{Illinois}
  \country{USA}
}
\email{ogokdemir@uchicago.edu}

\author{Samantha Grimes}
\orcid{0000-0003-1124-2956}
\affiliation{%
  \institution{University of Michigan}
  \city{Ann Arbor}
  \state{Michigan}
  \country{USA}
}
\email{samang@umich.edu}

\author{Adam Lauring}
\orcid{0000-0003-2906-8335}
\affiliation{%
  \institution{University of Michigan}
  \city{Ann Arbor}
  \state{Michigan}
  \country{USA}
}
\email{alauring@med.umich.edu}

\author{Brian R. Wasik}
\orcid{0000-0001-5442-3883}
\affiliation{%
  \institution{Cornell University}
  \city{Ithaca}
  \state{New York}
  \country{USA}
}
\email{brw72@cornell.edu}

\author{Anuj Karpatne}
\authornote{Corresponding authors.}
\orcid{0000-0003-1647-3534}
\affiliation{%
  \institution{Virginia Tech}
  \city{Blacksburg}
  \state{Virginia}
  \country{USA}
}
\email{karpatne@vt.edu}

\author{T. M. Murali}
\authornotemark[1]
\orcid{0000-0003-3688-4672}
\affiliation{%
  \institution{Virginia Tech}
  \city{Blacksburg}
  \state{Virginia}
  \country{USA}
}
\email{murali@cs.vt.edu}

\renewcommand{\shortauthors}{Antony et al.}

\begin{abstract}
    The lack of high-quality ground truth datasets to train machine learning (ML) models impedes the potential of artificial intelligence for science (AI for science) research.
    Scientific information extraction (SIE) from the literature using LLMs is emerging as a powerful approach to automate the creation of these datasets.
    However, existing LLM-based approaches and benchmarking studies for SIE focus on broad topics such as biomedicine and chemistry, are limited to choice-based tasks, and focus on extracting information from short and well-formatted text.
    The potential of SIE methods in complex, open-ended tasks is considerably under-explored.
    In this study, we used a domain that has been virtually ignored in SIE, namely virology, to address these research gaps.
    We design a unique, open-ended SIE task of extracting mutations in a given virus that modify its interaction with the host.
    We develop a new, multi-step retrieval augmented generation (RAG) framework called \ourframework for SIE. 
    In parallel, we curate a novel dataset of $629$ mutations in ten influenza A virus proteins obtained from $239$ scientific publications to serve as ground truth for the mutation extraction task.
    Finally, we demonstrate \ourframework{}'s superior performance using a novel and comprehensive evaluation and comparison with vanilla RAG and other state-of-the art RAG- and
    agent-based tools 
    for SIE.
\end{abstract}


\begin{teaserfigure}
  \centering
  \includegraphics[width=.9\textwidth]{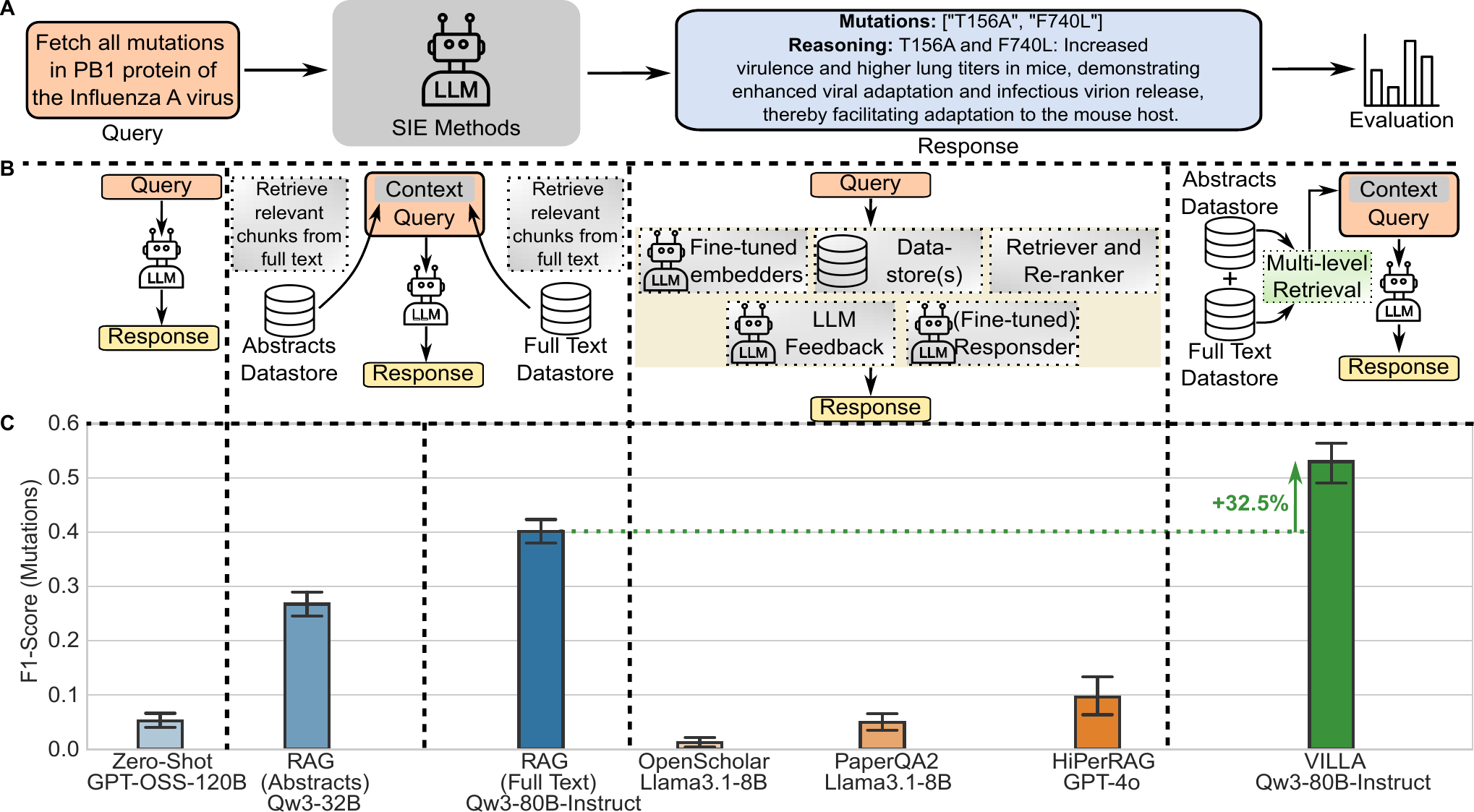}
  \caption{(A) We study an open-ended task in scientific information extraction (SIE) in the context of virology. (B) We develop \ourframework, a novel two-stage method for retrieval-augmented generation (RAG). (C) \ourframework outperforms zero-shot prompting, RAG- and agent-based baselines, and other state-of-the-art methods for SIE in our novel task of viral mutation extraction.}
  \Description{}
  \label{fig:teaser}
\end{teaserfigure}


\maketitle
\section{Introduction}
\label{sec:introduction}
Artificial Intelligence (AI) for science is accelerating research through discovery of patterns and candidate hypotheses from large-scale datasets.
However, a major barrier in AI for Science is the lack of high-quality ground truth datasets to train machine learning (ML) models~\cite{berens-montgomery-ai4science-arxiv-2023, ai4science-natureresearcjhintell-2025}.
In many fields, the relevant information is spread across the literature as unstructured text~\citep{holmes-goodrum-virology-landscape-cell-2024, gokdemir-ramanathan-hiperrag-pascconf-2025}.
It is infeasible to manually review relevant publications~\citep{hanson-brockington-strain-scientific-publishing-mitquantitivesciencestudies-2024, brack-ewerth-sie-domains-advancesinforetrieval-2020, dagdelen-jain-sie-material-science-natcommunications-2024, ma-si-virology-research-21stcentry-virologysinica-2024}, especially with the exponential growth in scientific literature~\citep{fire-guestrin-increasing-publications-gigascience-2019, elshamieh-chebly-increasing-publications-bmcsystreviews-2024}.
Hence, there is an urgent need for automated mechanisms to extract the findings hidden in the scientific corpus to advance AI for science.

Scientific information extraction (SIE) is the process of identifying and retrieving the desired data from the unstructured text in research publications.
Advances in SIE focus on mining important data from text to  enable auto-generation of tabular databases of the extracted information~\citep{dagdelen-jain-sie-material-science-natcommunications-2024}.
In recent years, the promising potential of LLMs in understanding general language and reasoning tasks have led to the development of LLM-powered tools for SIE~\citep{xu-chen-llm-generative-ie-frontiersofcs-2024}.
Various approaches adopt domain specific LLMs~\citep{phan-altan-bonnet-scifive-arxiv-2021, gu-poon-pubmedbert-blurb-acmtranscomputhealthcare-2021, luo-liu-biogpt-briefbioinform-2022, jin-lu-medcpt-bioinformatics-2023} and general purpose LLMs for SIE~\citep{garcia-papa-sie-arxiv-2024, chen-xu-benchmark-llm-biomedical-natcommun-2025, ateia-almohaishi-llm-sie-business-process-mgmnt-trendstheorypracticedigitallibraries-2025, shang-hou-geneturing-briefbioinform-2025} using zero-shot and few-shot prompting, retrieval-augmented generation (RAG), and supervised fine-tuning (SFT) (please see \suppnote{\ref{appndx_sec:relatedwork}} for brief descriptions).
Multiple studies~\citep{jimenezgutierrez-su-sie-benchmark-gpt3-icl-aclemnlp-2022, shamsabadi-auer-sie-virology-covid19-ner-eacl-2024, chen-xu-benchmark-llm-biomedical-natcommun-2025, shang-hou-geneturing-briefbioinform-2025} demonstrate and compare the ability of open and closed LLM-driven tools to mine information from abstracts and full text of scientific publications.
\newtext{Despite the success of these methods, there are two issues.}

   \paragraph{(1) Standard information extraction tasks are inadequate in capturing the complexity of SIE}
   Existing studies evaluate the capabilities of LLMs in SIE through tasks such as named entity recognition (NER)~\citep{jimenezgutierrez-su-sie-benchmark-gpt3-icl-aclemnlp-2022, shamsabadi-auer-sie-virology-covid19-ner-eacl-2024, chen-xu-benchmark-llm-biomedical-natcommun-2025}, relation extraction (RE)~\citep{jimenezgutierrez-su-sie-benchmark-gpt3-icl-aclemnlp-2022, chen-xu-benchmark-llm-biomedical-natcommun-2025}, question answering (QA), and document classification (DC).
    All these tasks except NER are formulated in a manner that includes the desired information in the prompt.
    QA explicitly provides two or more (one correct and remaining incorrect) options. IR, RE, and DC implicitly include a list of possible labels in the prompt for the LLMs to choose from.
    Moreover, all these tasks involve extracting information from short, clean, and well-formatted scientific text, such as a sentence or an abstract.
    These tasks do not realize the complexity of parsing the full text of publications and extracting the correct information.

     \paragraph{(2) The potential of SIE methods in complex, open-ended tasks is unknown}
     Prior benchmarking pipelines evaluate the ability of several general purpose LLMs in SIE using zero-shot~\citep{chen-xu-benchmark-llm-biomedical-natcommun-2025} and few-shot prompting~\citep{jimenezgutierrez-su-sie-benchmark-gpt3-icl-aclemnlp-2022, chen-xu-benchmark-llm-biomedical-natcommun-2025}. 
    Additionally, many studies report and compare the performance of LLMs fine-tuned for SIE~\citep{jimenezgutierrez-su-sie-benchmark-gpt3-icl-aclemnlp-2022, chen-xu-benchmark-llm-biomedical-natcommun-2025}.
    Though the fine-tuning approach yields the best performance, the compute resources requirement for SFT of LLMs is high~\citep{jimenezgutierrez-su-sie-benchmark-gpt3-icl-aclemnlp-2022, chen-xu-benchmark-llm-biomedical-natcommun-2025}.
    Several RAG-based tools such as HiPerRAG~\cite{gokdemir-ramanathan-hiperrag-pascconf-2025} and agentic frameworks such as OpenScholar~\citep{asai-hajishirzi-openscholar-arxiv-2024}, and PaperQA2~\citep{skarlinski-white-paperqa2-arxiv-2024} that exploit the advantages of RAG and SFT~\citep{asai-hajishirzi-openscholar-arxiv-2024} have demonstrated superior performance in SIE.
    However, the evaluation of these methods was limited to choice-based tasks.
    There are no studies that evaluate and compare the performance of RAG-based methods in open-ended tasks.

In this study, we have addressed both these major research gaps. We used a highly under-explored domain in SIE, namely virology, to evaluate and demonstrate the potential of our contributions.
Akin to other domains, the literature in virology has rapidly expanded (particularly since the COVID-19 pandemic) from 40,000 papers annually in 2019 to over 80,000 in subsequent years.
It is imperative to extract pertinent information from this vast corpus using SIE methods and convert them into suitable formats for computational analysis to accelerate research in virology.
However, there are few ground-truth datasets for SIE tasks in virology due to the paucity of annotated datasets in this domain.
The standard datasets used to benchmark SIE methods span a wide range of topics such as drugs~\citep{crichton-korhonen-bc5chem-bc5disease-datasets-bmcbioinformatics-2017, nye-wallace-ebmpico-dataset-acl-2018, islamajdogan-lu-chemprot-dataset-databaseoxford-2019, herrero-zazo-declerck-ddi-dataset-biomedicalinformatics-2013}, diseases~\citep{dogan-lu-ncbi-disease-dataset-biomedicalinformatics-2014, hanahan-weinberg-hoc-dataset-cell-2011, chen-lu-litcovid-dataset-nar-2021}, molecular biology~\citep{smith-wilbur-bc2gm-dataset-genomebiology-2008, collier-kim-jnlbpa-dataset-nlpbabionlp-2004, bravo-furlong-gad-dataset-bmcbioinformatics-2015}, and general biomedical concepts~\citep{nentidis-kakadiaris-bioasq-bionlpworkshopacl-2017, jin-lu-pubmedqa-dataset-emnlp-ijcnlp-2019}. See \suppnote{\ref{appndx_sec:benchmark-datasets}} for details on the datasets.
Barring LitCovid~\citep{chen-lu-litcovid-dataset-nar-2021}, the proportion of questions pertaining virology in these datasets is low ($0.34\%$ to $5.32\%$).
LitCovid is a virology dataset focusing only on SARS-CoV-2.

\todoblessy{Add figure descriptions}

In this study, 
\begin{enumerate*}[(i)]
    \item we designed an open-ended SIE task, i.e., the prompt we created did not ask LLMs to select from a set of pre-defined labels or answers. This task prompted the LLM to extract mutations in a given viral protein. The query did not include any concise or formatted text containing the potential answer.
    \item We proposed a novel framework called `

    \ourframework' for SIE. This two-stage RAG-based framework includes a new technique to construct the context by using abstracts to identify relevant papers and then performing an exhaustive search using the full-text of the selected publications.
    \item Our team of domain experts (virologists) manually curated a novel dataset in virology to address the lack of annotated ground truth in this domain. This dataset contained $629$ mutations across ten influenza A virus proteins from $239$ scientific publications to benchmark SIE methods.
    \item We demonstrated the superior performance of \ourframework using a comprehensive evaluation and comparison with vanilla RAG and three other state-of-the art (SOTA) RAG- and agent-based tools involving both open and closed LLMs to extract information from scientific publications. We leveraged our expert-curated dataset as the ground truth for these evaluations.
\end{enumerate*}
Though we have focused on virology in this study, our proposed two-stage method \ourframework can be readily adapted to other domains.
Our benchmarking strategy for RAG-based tools is also generalizable, contingent on the availability of a set of papers containing the desired information and a ground-truth dataset to evaluate the findings.

\section{Methodology}
\label{sec:methodology}

\subsection{Problem statement} \label{methods:viral-mutation-extraction-task}
We defined an open-ended task called `viral mutation extraction' to evaluate the ability of LLMs to extract information from scientific publications~(\fig{\ref{fig:teaser}A}).
The goal of the task was to retrieve mutations in a given viral protein that impact virus-host interaction from one or more abstracts or full-text publications.
The input included the name of a virus and one of its protein.
The expected output of the task was a list of mutations in the given viral protein along with the effect of each mutation in the virus-host interaction.

With guidance from virologists, we developed a prompt template to query LLMs.
In the prompt, we described the viral mutations extraction task, and provided the name of a virus and a protein of interest.
The prompt also contained instructions for the LLM to respond with a JSON output containing two fields namely `mutations' and `reasoning' capturing the list of mutations and the effects of these mutations on virus-host interaction respectively.

We used this prompt template to query several LLMs using zero-shot prompting and SOTA SIE tools such as HiPerRAG, OpenScholar, and PaperQA2~(\suppfig{\ref{appndx_fig:prompt-zero-shot}}).
We adapted the same template for three RAG-based methods designed in this study namely RAG with abstracts, RAG with full text, and \ourframework~(\suppfig{\ref{appndx_fig:prompt-rag}}). 
Specifically, we added instructions for the LLMs to retrieve the correct mutations from only within the contextual information provided in the prompt.

\subsection{Background} \label{methods:rag}
Our proposed framework for SIE is based on an AI technique called retrieval augmented generation~(RAG).
The RAG method enhances the knowledge captured in the parameters of LLMs at inference time using information from external sources.
It is an optimal method of improving the performance of general purpose LLMs for domain specific tasks without incurring the overhead of compute resources required in supervised fine-tuning.

In this method, we supplement the query in the prompt with a `context' that contains the correct answer. This context is constructed using relevant information from a datastore of documents of any type. The prompt also instructs the LLMs to look for the answer only in the provided context.

A RAG framework~\cite{lewis-kiela-rag-neurips-2020}  typically consists of four components namely, 
\begin{enumerate*}[(i)]
    \item datastore ($\mathcal{D}$),
    \item embedder ($\mathcal{E}$),
    \item retriever ($\mathcal{R}$), and
    \item response generator ($\mathcal{G}$).
\end{enumerate*}
The working of RAG with respect to the SIE experiments designed in this study is as follows:
\begin{enumerate}
    \item In SIE, scientific articles act as documents to be store in the datastore.
    \item We divided the text from publications into $n$ chunks ($c_1, c_2, \dots, c_n$) of size $s$ each.
    \item `Embedder' $\mathcal{E}$ (we used an LLM in this study) encoded each chunk $c_i$ into embedding vectors $\mathbf{c_i}$. We stored these $n$ embeddings in a vector `datastore' $\mathcal{D}$ which served as the knowledge store for the RAG framework.
    \item At inference time, $\mathcal{E}$ encoded the prompt to the LLM into a vector $\mathbf{p}$.
    \item The `retriever' $\mathcal{R}$ computed distances between $\mathbf{p}$ and all chunk vectors $\mathbf{c_i}$ in $\mathcal{D}$ using cosine similarity as defined in equation \ref{eq:cosine-distance}. $\mathcal{R}$ selected the top $k$ chunk embeddings that were closest in distance to $\mathbf{p}$ AND whose distance was below a set threshold $t$. The retriever then returned the text corresponding to each of the $k$ selected chunks. This selected text formed the context $C$ containing external knowledge from scientific publications.
        \begin{equation} \label{eq:cosine-distance}
            d(\mathbf{p}, \mathbf{c_i}) = 1 - \frac{\mathbf{p} \cdot \mathbf{c_i}}{\|p\|\|\mathbf{c_i}\|}
        \end{equation}
    \item We concatenated the selected chunks in order of increasing distances to form the context $C$. We inserted this context into the original prompt.
    \item We passed the augmented query to the response generator LLM $\mathcal{G}$ which generated a response for the prompt. We expected $\mathcal{G}$ to extract the desired information from the context only.
\end{enumerate}


\subsection{Baselines}
\label{sec:baselines}
We compared the performance of our proposed framework with three baselines: zero-shot prompting and two types of RAG implementations.
We summarize each method below with details in \suppnote{\ref{appndx_sec:baselines}}.

\subsubsection{Zero-shot prompting} \label{methods:zero-shot-prompt}
We evaluated eleven general purpose LLMs using zero-shot prompting for viral mutation extraction~(\sctn{\ref{methods:viral-mutation-extraction-task}}).
Here, we did not augment the prompt with any additional information.
Therefore, the LLMs responded using the parametric knowledge learned during their training.

\subsubsection{RAG with abstracts} \label{methods:rag-abstracts}
The datastore ($\mathcal{D_A}$) in this method contained embeddings of the abstracts of all the $239$ scientific publications in the influenza A dataset (\sctn{\ref{experiments:influenza-a-dataset}}).

%

\subsubsection{RAG with full text} \label{methods:rag-full-text}
The datastore ($\mathcal{D_F}$) in this method contained embeddings of $10,327$ chunks of the textual information from $239$ scientific publications in the influenza A dataset (\sctn{\ref{experiments:influenza-a-dataset}}).

We measured the performance of seven different LLMs as embedders encoding the abstracts or full text of publications (\supptab{\ref{appndx_tab:embedder-llms}}). Similarly, we evaluated eleven distinct LLMs as responders that extracted mutation information from the context provided in the prompts.

\subsection{Proposed framework: \ourframework} \label{methods:multi-level-rag}
\begin{figure*}[ht]
    \centering
    \includegraphics[width=\linewidth]{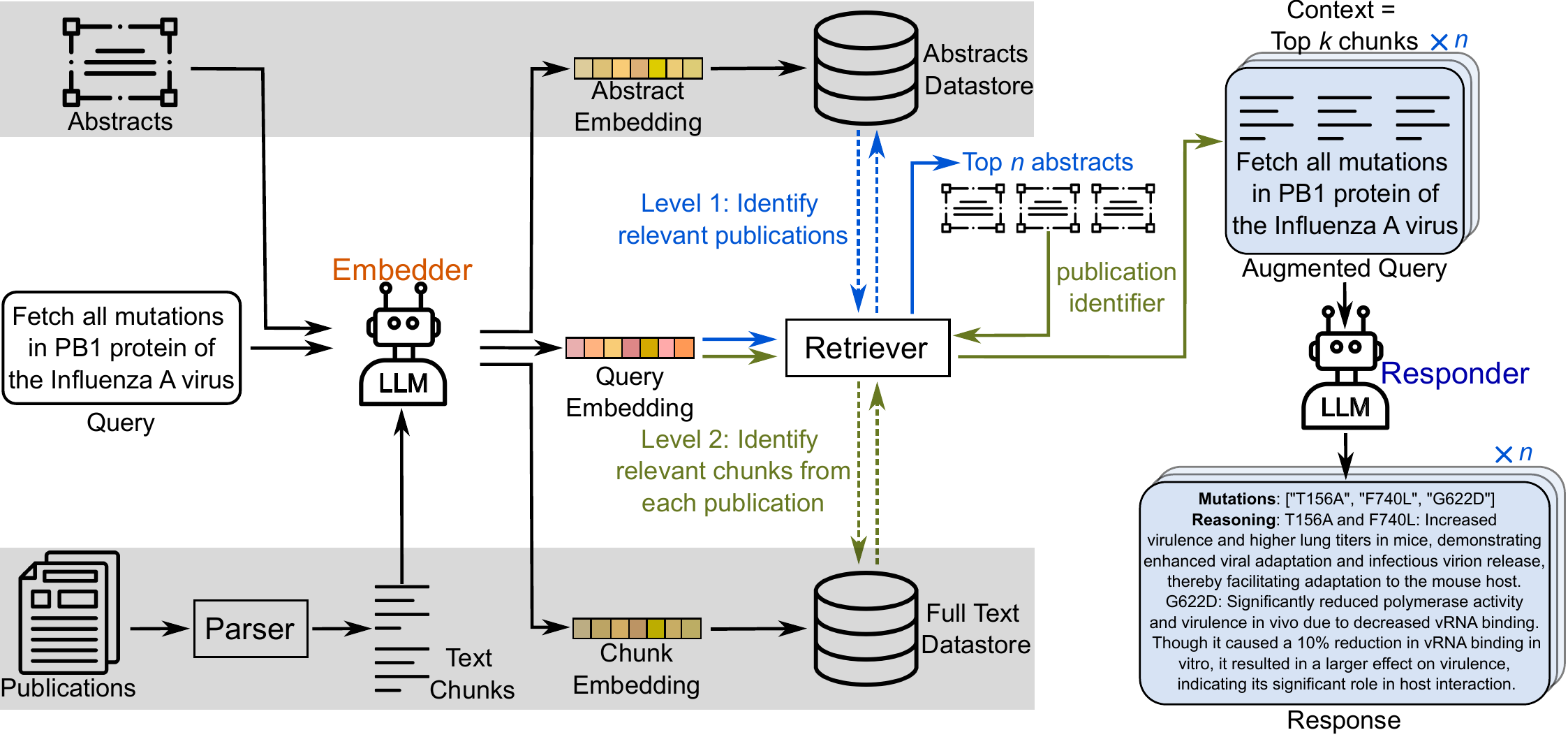}
    \caption{Overview of the \mlr framework called `\ourframework' for SIE from publications. For a given prompt, the framework first identifies relevant publications using embeddings in the abstract datastore. Then for each selected publication, (i) retrieves relevant chunks of full text from the full text datastore, 
    (ii) augments the input prompt with the concatenation of all the selected chunks, and
    (iii) generates a response for the augmented prompt.}
    \Description{}
    \label{fig:villa-architecture}
\end{figure*}

We developed \ourframework to combine the power of RAG with abstracts and RAG with full text in two steps (Algorithm \ref{alg:mlr} and \fig{\ref{fig:villa-architecture}}):
\begin{enumerate*}[(i)]
    \item identify the relevant publications, and
    \item retrieve the accurate information from the selected publications.
\end{enumerate*}
The different components of the method included datastore for abstracts $\mathcal{D_A}$, datastore for chunks of textual information in publications $\mathcal{D_F}$, embedder $\mathcal{E}$, retriever $\mathcal{R}$, and a response generator $\mathcal{G}$ (responder).
It also required a corpus of scientific publications with abstracts and full text, and a prompt $p$ that described the information to be retrieved from the corpus.
The variable parameters in the method were $k_a$: number of abstracts to be retrieved, $k_c$: number of chunks to be retrieved from each publication, and a distance threshold $t$. 

\algrenewcommand\algorithmicrequire{\textbf{Inputs:}}
\algrenewcommand\algorithmicensure{\textbf{Initialize:}}
\begin{algorithm}
    \caption{\ourframework} \label{alg:mlr}
    \begin{algorithmic}[1]
    \Require corpus of scientific publications, $\mathcal{D_A}$, $\mathcal{D_F}$, $\mathcal{E}$, $\mathcal{R}$, $\mathcal{G}$, $p$, $k_a$, $k_c$, and $t$.
    \Ensure Initialize the datastores $\mathcal{D_A} \text{ and } \mathcal{D_F}$.
    \For{each publication in the corpus}
        \State Embed the abstract using $\mathcal{E}$ and store the abstract embedding with the corresponding publication identifier in $\mathcal{D_A}$.
        \State Break the full text of the publication into chunks of equal size with overlap.
        \State Embed each chunk using $\mathcal{E}$ and store the chunk embedding with the corresponding publication identifier in $\mathcal{D_F}$.
    \EndFor

    \Function{VILLA}{$\mathcal{D_A}, \mathcal{D_F}, \mathcal{E}, \mathcal{R}, \mathcal{G}, k_a, k_c, p$}
    \State Embed the prompt $p$ using $\mathcal{E}$.
    \State Level 1 retrieval: Retrieve the $k_a$ abstracts from $\mathcal{D_A}$ with the closest cosine distance (within threshold $t$) to the prompt embedding.
    \For{each of the selected abstracts}
        \State Level 2 retrieval: Retrieve $k_c$ chunks of publication, corresponding to the selected abstract, from $\mathcal{D_F}$ with the closest cosine distance (within threshold $t$) to the prompt embedding. These chunks form the context.
        \State Concatenate the retrieved context with the original prompt $p$ to augment it.
        \State Query the responder $\mathcal{G}$ with the augmented prompt.
        \State Store the mutations and reasoning from the response of $\mathcal{G}$.
    \EndFor
    \State \Return Lists of mutations and reasonings from LLMs for all selected publications.
    \EndFunction
    \end{algorithmic}
\end{algorithm}

\subsection{Expert curated influenza A dataset} \label{experiments:influenza-a-dataset}
\begin{figure}[ht]
    \centering
    \includegraphics[width=\linewidth]{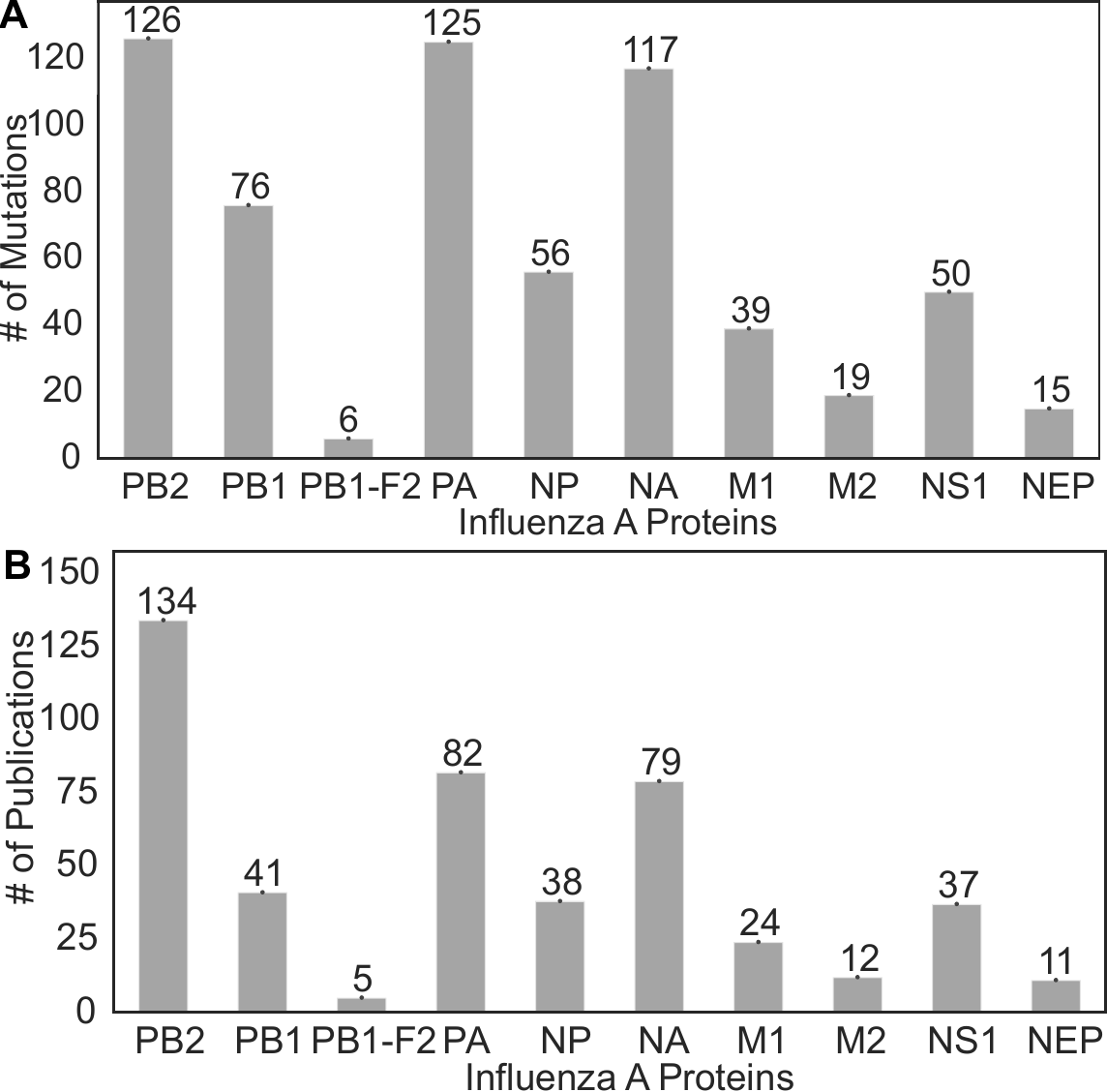}
    \caption{The number of \textbf{(B)} publications and \textbf{(A)} mutations with findings about mutations impacting virus-host interaction, in ten influenza A proteins.}
    \Description{}
    \label{fig:influenza-a-dataset}
\end{figure}
To evaluate the ability of different LLM-driven SIE methods to extract foundational knowledge in virology, we focused on influenza A virus. Influenza A viruses are extremely successful avian-borne viruses that frequently spill-over and emerge into mammalian hosts to cause outbreaks and epidemics. Influenza has been the source of novel human pandemics in 1918 (H1N1), 1957 (H2N2), 1968 (H3N2), and 2009 (swine-origin H1N1), while also contributing to millions of cases of severe illness globally each year during  seasonal circulation~\citep{krammer-garcia-sastre-influenza-mutations-natrevdisprimers-2018}. For both pandemic human influenza and emergent lineages of virus in non-human mammals (i.e. H5N1), a multitude of studies have identified specific mutations in influenza viral proteins that demonstrate (or infer) increased risk of mammalian adaptation, virulence, and transmission to facilitate pandemic potential.

We manually curated a dataset of mutations in the influenza A virus genome from primary literature sources.
The dataset focused on observed and experimental mutations with identified risk potential for mammalian adaptation.
We coded all mutations for the specific amino acid change (alpha-numeric-alpha) and the viral protein as described in the text.
We excluded mutations in the hemagglutinin (HA) protein as amino acid numbering schematics in the literature were inconsistent and often transformed from alignments and subtype designations~\citep{burke-smith-influenza-a-dataset-plosone-2014}.
There were a total of ten viral proteins categorized by major open reading frames (PB2, PB1, PA, NP, NA, M1, NS1) and known leaky ribosomal scanning or splice variants (PB1-F2, M2, NEP).

This dataset contained $629$ mutations in ten influenza A virus proteins obtained from $239$ unique scientific publications (\fig{~\ref{fig:influenza-a-dataset}}).

\subsection{Evaluation strategy}\label{methods:quantitative-evaluation}
In this study, we used the influenza A dataset as ground truth~(\sctn{\ref{experiments:influenza-a-dataset}}) to evaluate the different SIE methods. Each prompt to the LLMs included one of the ten different viral proteins in the ground truth dataset. We formulated two different types of evaluations:

\subsubsection{Evaluation of retrieved context} \label{methods:evaluation-context}
We evaluated the embedder LLMs in the RAG with abstracts and full text, and the proposed \ourframework methods (\suppnote{\ref{appndx_sec:llms-embedders}}, \supptab{\ref{appndx_tab:embedder-llms}}).
The ground truth dataset~(\sctn{\ref{experiments:influenza-a-dataset}}) included lists of relevant publications for each viral protein.
We used these lists to gauge the correctness of the context.
For each prompt that contained a viral protein, we captured the publication corresponding to each of the top $k$ abstracts (RAG with abstracts and \ourframework) or chunks (RAG with full text) that composed the context.
We compared this list of retrieved publications with those corresponding to the input protein in the ground truth dataset.
We computed the precision, recall, and F1-score of the retrieved context for each protein (prompt).

The retriever selected the top $k$ chunks closest to the embedding of the prompt. Thus, this evaluation of the context helped to compare the different embedder LLMs.

\subsubsection{Evaluation of extracted information} \label{methods:evaluation-information}
Our ground truth dataset consisted of a list of mutations impacting virus-host interaction for every viral protein.
In the SIE task, we prompted the responder models~(\suppnote{\ref{appndx_sec:llms-generators}}, \supptab{\ref{appndx_tab:generator-llms}}) to respond with a list of mutations in given viral protein.
Further we defined the symbolic representation of mutations as $<$original amino acid$><$position$><$changed amino acid$>$ where $<$position$>$ denoted the position where the mutation occurs in the viral protein sequence, $<$original amino acid$>$ is the alphabet representing the amino acid at that position in the viral protein sequence, and $<$changed amino acid$>$ is the alphabet representing the changed amino acid. For example, ``A123C'' is a mutation of amino acid ``A'' to ``C'' at position ``123''.
We compared the list of mutations in the response of the LLMs with the mutations corresponding to the input protein in the ground truth dataset~(\sctn{\ref{experiments:influenza-a-dataset}}).
We computed the precision, recall, and F1-score of the retrieved mutations for each protein (prompt).

\subsubsection{Qualitative evaluation} \label{methods:qualitative-evaluation}
Two virologists who earned their doctoral degrees in virology evaluated the `reasoning' component in the LLM responses.
We designed a five-point rubric (\suppnote{\ref{appndx_sec:qualitative_evaluation}}) in five categories for the qualitative evaluation of LLM responses, with one being the lowest score and five the highest .
Rubric categories were clarity, conciseness, correctness, citations/context, and contribution. 
\newtext{We developed a web-based user interface where the domain experts could view and evaluate the responses in an unbiased manner (\suppnote{\ref{appndx_sec:user-interface-qualitative-evaluation}})}.
We defined each category as a question for the evaluator in order to further standardize the qualitative considerations.

\section{Experimental Setup}
We evaluated a wide range of LLMs as response generators using zero-shot prompting and RAG.
In addition to the generators, we also evaluated different embedding LLMs as embedders in RAG.
Finally we benchmarked three state-of-the-art RAG- and agent-based tools for SIE. 
\subsection{Response generators for zero-shot prompting and RAG-based methods}
We evaluated eleven general purpose LLMs using the SIE task~(\sctn{\ref{methods:viral-mutation-extraction-task}}) and the influenza A dataset~(\sctn{\ref{experiments:influenza-a-dataset}}).
We used two closed models namely OpenAI's o1 and GPT-4o, and nine open models including GPT-OSS 120B, ZAI-GLM4.5, two models from the DeepSeek family, two from Llama, and three from Qwen3 (\suppnote{\ref{appndx_sec:llms-generators}}, \supptab{\ref{appndx_tab:generator-llms}}).

\subsection{Embedders for RAG-based methods}
The three RAG-based frameworks (Sections \ref{methods:rag-abstracts}, \ref{methods:rag-full-text}, \ref{methods:multi-level-rag}) contained an additional LLM-based component called embedder~(\sctn{\ref{methods:rag}}).
We compared the performance of seven different textual embedders that encoded text input into vectors.
\subsubsection{General purpose embedders.} We used one closed namely Open AI's text-embedding-3-large, and four open embedders including Llama 3.1:8B, Qwen3-Embedding:8B, Google's Embedding Gemma:300M, and Embedding Mistral-Q4KM from SalesForce~(\suppnote{\ref{appndx_sec:llms-embedders}}).
The proportion of scientific content is low in the training dataset of a typical LLM. General-purpose encoding LLMs underperform in tasks involving scientific text~\cite{gokdemir-ramanathan-hiperrag-pascconf-2025}.
\subsubsection{Embedders for scientific publications}
We also evaluated two additional embedder LLMs pretrained explicitly on textual information from scientific publications.
PubMedBERT is a Bidirectional Encoder Representations from Transformers (BERT) model pretrained on 14 million abstracts from PubMed~\citep{gu-poon-pubmedbert-blurb-acmtranscomputhealthcare-2021}.
MedCPT consists of query encoder and document encoder which are two independent PubMedBERT models. These two encoders were further pretrained using contrastive learning on $255$ million query-article pairs from PubMed search logs. During this constrastive learning, the model learned to maximize the similarity between a query and its related article(s).~\citep{jin-lu-medcpt-bioinformatics-2023}.

\subsection{Other frameworks} \label{methods:sota-frameworks}
We evaluated three state-of-the-art methods for SIE: 
\begin{enumerate}[(i)]
    \item OpenScholar~\citep{asai-hajishirzi-openscholar-arxiv-2024}: enhanced RAG framework with a large-scale scientific datastore consisting of $45$ million open-access papers;
    \item PaperQA2~\citep{skarlinski-white-paperqa2-arxiv-2024}: a scientific question-answering framework that utilises an agentic framework; and
    \item HiPerRAG~\cite{gokdemir-ramanathan-hiperrag-pascconf-2025}: a modular, open-source high-performance computing (HPC) framework for scaling RAG across the full scientific pipeline.
\end{enumerate}
Please see \suppnote{\ref{appndx_sec:baselines}} for details.

We compared the performances of all frameworks with Llama 3.1:8B as the response generator (responder) in a RAG-based setup.
We also assessed additional flavors of PaperQA2, HiPerRAG, and \ourframework with different LLMs as responders and varying compositions of the datastores in the respective frameworks.
OpenScholar~\citep{asai-hajishirzi-openscholar-arxiv-2024} is powered through a fine-tuned Llama 3.1:8B model and a datastore of $45$ million open-access scientific papers.
In PaperQA2, we leveraged the capabilities of the full agentic framework with Llama3.1:70B as the agent and Llama3.1:8B as the query model (referred to as `Agents' mode in \tab{\ref{tab:villa-baseline-comparison}}). We also evaluated an ablation of PaperQA2 comprising only the RAG capabilities (`RAG' mode).
PaperQA2 enables users to build a custom datastore to perform SIE and we used the $239$ influenza A publications in order to mimic the setup in \ourframework.
For HiPerRAG \cite{gokdemir-ramanathan-hiperrag-pascconf-2025}, we implemented a high-performance RAG framework using $\sim440\text{K}$ full-text microbiology publications to construct the datastore. 
We also created an additional datastore of the $239$ influenza A articles and evaluated HiPerRAG with this smaller set of papers.
\section{Results}
\label{sec:results}
We prompted LLMs to respond with the mutations in a given viral protein.
An experiment in this study included ten such prompts, one for each influenza A protein~(\sctn{\ref{experiments:influenza-a-dataset}}).
We repeated each experiment five times to ensure reliability of the LLM responses. 
For each method, we report the distribution of metrics across these fifty experiments~(\sctn{\ref{methods:quantitative-evaluation}}).

\subsection{Performance of zero-shot prompting} \label{results-zero-shot-prompting}
\begin{figure}[ht]
    \centering
    \includegraphics[width=\linewidth]{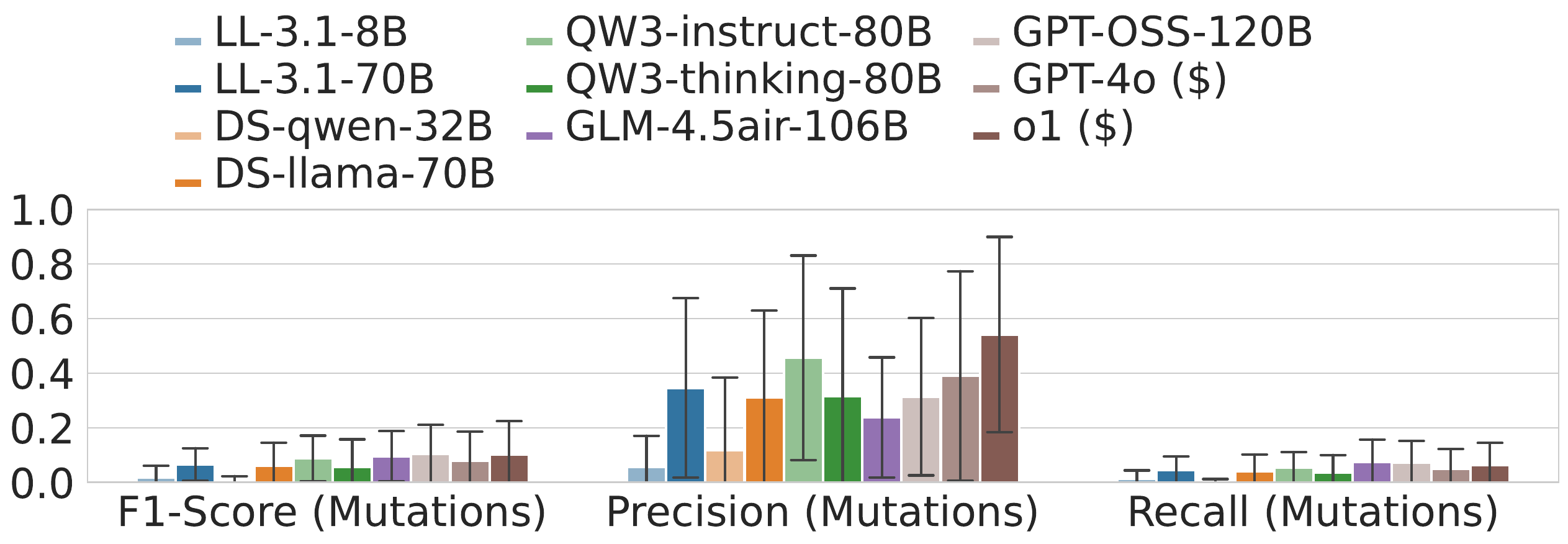}
    \caption[Evaluation of LLMs for viral mutation extraction using zero-shot prompting.]{Distribution of the  F1-scores, precision, and recall of eleven LLMs in identifying mutations across ten proteins in influenza A virus using zero-shot prompting. Each bar height and error bar corresponds to the mean and standard deviation of the distribution of scores across ten proteins and five iterations, respectively. Abbreviations: LL, Llama; DS, DeepSeek; QW3, Qwen.}
    \Description{}
    \label{fig:results-zero-shot-prompting-metrics}
\end{figure}

We evaluated eleven LLMs in SIE using zero-shot prompting for viral mutation extraction in influenza A virus~(\suppnote{\ref{appndx_sec:results-zero-shot-prompting}}).
For each model, we averaged the precision, recall, and F1-scores over all proteins.
The Open AI open source model GPT-oss-120B had the highest average F1-score of $0.10\pm0.11$ across all proteins and iterations, followed by Open AI o1 (\$) with $0.10\pm0.12$ (\fig{\ref{fig:results-zero-shot-prompting-metrics}}). 
However, the average recall was low, ranging from $0.003$ (Deepseek-R1-Distill-Qwen-32B) to $0.07$ (Z.ai-GLM4.5-Air).
On average, the models were able to accurately retrieve only $4.26\%$ of the mutations across all proteins.
We concluded that zero-shot prompting was not an effective approach for retrieving influenza A viral mutations.
General purpose LLMs are not exposed to domain specific knowledge in scientific publications during pretraining~\citep{gokdemir-ramanathan-hiperrag-pascconf-2025}.
The lack of information about virology in the parametric knowledge of the LLMs motivated us to explore RAG-based methods for SIE.

\subsection{Viral mutation extraction using RAG with abstracts} \label{results-rag-abstracts}
Next, we set up a RAG framework using abstracts~(\sctn{\ref{methods:rag-abstracts}}) of all $239$ scientific publications in the influenza A dataset~(\sctn{\ref{experiments:influenza-a-dataset}}).

\begin{figure*}[ht]
    \centering
    \includegraphics[width=\textwidth]{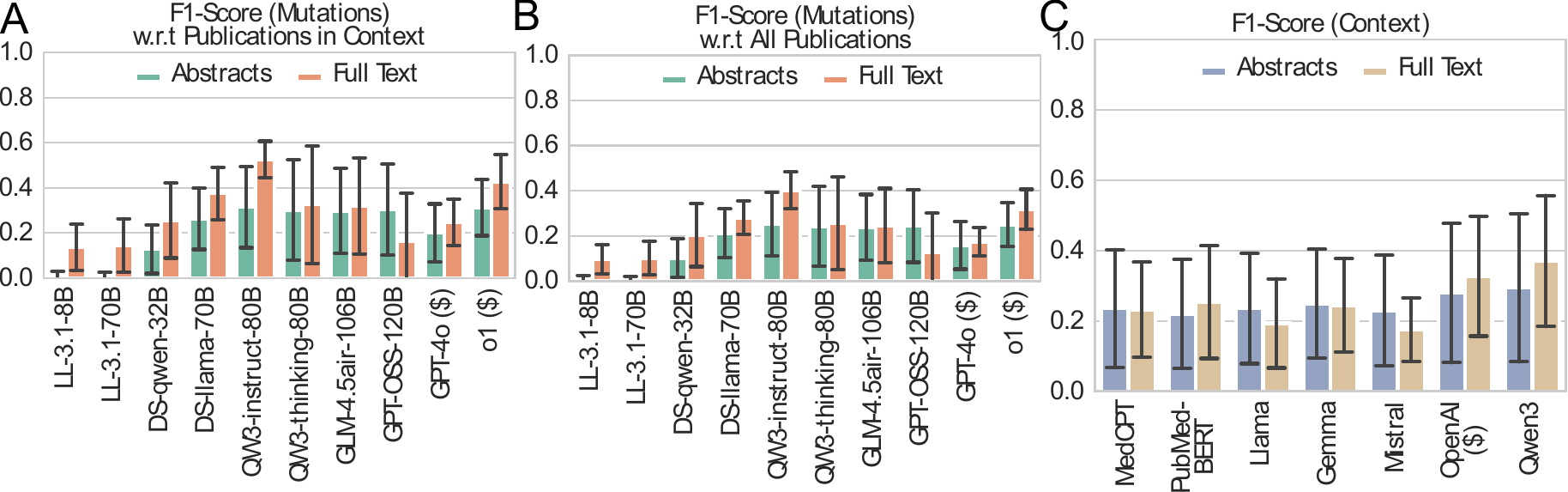}
    \caption[Evaluation of vanilla RAG frameworks for viral mutation extraction.]{Distribution of the F1-scores with respect to the \textbf{(A)} ground truth mutations in the publications represented in the context in the prompt and \textbf{(B)} the ground truth mutations in all the publications. 
    \textbf{(C)} Distribution of the F1-scores of the retriever in the two RAG frameworks in identifying abstracts relevant to each of the ten proteins in influenza A virus. The retriever is evaluated using seven different embedding LLMs. Each bar height and error bar corresponds to the mean and standard deviation of the distribution of scores across ten proteins and five iterations, respectively. Abbreviations: LL, Llama; DS, DeepSeek; QW3, Qwen.}
    \Description{}
    \label{fig:results-vanilla-rag-f1scores}
\end{figure*}

\subsubsection{Evaluation of the extracted mutations} \label{results-rag-abstracts-mutations}
We evaluated the retrieved mutations at two levels:
\begin{enumerate*}[(i)]
    \item with respect to the mutations for each protein in the ground truth dataset, and
    \item with respect to context, that is, with respect to only those ground truth mutations present in the publications of the abstracts selected as part of the the context.
\end{enumerate*}
Note that for each protein, the latter set of mutations was a subset of the former.

Qwen3-32B had the overall best performance with average precision, recall, and F1-scores of $0.82 \pm 0.13$, $0.16 \pm 0.06$, and $0.27 \pm 0.08$ respectively~(\fig{\ref{fig:results-vanilla-rag-f1scores}}, \suppfig{\ref{appndx_fig:results-vanilla-rag-all-precision-recall}}). Irrespective of the precision, we observed low recall and F1-scores across all models suggesting the abstracts in these publications did not contain the mutations we were seeking. 

For all responder LLMs, the precision with respect to context and overall precision scores were strictly identical with the exception of one protein (NA) where the overall precision was larger in some iterations. The recall with respect to context was larger than the overall recall scores.
This result suggested that the responder did not hallucinate correct mutations that were not present within the context.
The distribution of all scores for each of the ten influenza A proteins are in Appendix Figures~\ref{appndx_fig:results-rag-abstracts-mutations-all-proteins-metrics} and \ref{appndx_fig:results-rag-abstracts-mutations-wrt-context-all-proteins-metrics}.


\subsubsection{Evaluation of the selected abstracts} \label{results-rag-abstracts-context}
We evaluated the ability of this framework to retrieve the correct abstracts.
For a given protein in the prompt, we compared the publications represented in the context with those corresponding to the protein in the ground truth dataset.
The average F1-scores across all ten influenza A proteins are shown in \fig{\ref{fig:results-vanilla-rag-f1scores}C} while precision and recall scores are in \suppfig{\ref{appndx_fig:results-vanilla-rag-all-precision-recall}}. The distribution of all three metrics on a per protein basis is shown in \suppfig{\ref{appndx_fig:results-rag-abstracts-context-all-proteins-metrics}}.
The mean recall of the publications in the context across all proteins ranged from $0.54$ (MedCPT) to $0.70$ (Qwen3-Embedding:8B).
However, the range of mean precision was low: from $0.16$ (PubMedBERT) to $0.22$ (Qwen3-Embedding:8B).
Qwen3-Embedding:8B had the best performance with mean F1-score of $0.29\pm0.21$, followed by Open AI's text-embedding-3-large (\$) model with $0.28\pm0.20$.
We chose the Qwen3-Embedding:8B model with the highest F1-score for further analysis.

Next, we sought to understand the reason for the low precision.
We computed the cosine distances between the Qwen3-Embedding:8B model's embedding of a prompt for viral mutation extraction for a influenza A protein, and the embeddings of the abstracts from publications corresponding to the protein (relevant).
Similarly, we measured the distance between the prompt and abstracts of all remaining publications (non-relevant to the given protein) in the ground truth dataset.
We compared the distributions of distances of relevant and non-relevant abstracts for all ten influenza A proteins~(\suppfig{\ref{appndx_fig:cosine-distance-abstracts-qwen3-8b}}).
The mean distance of the relevant abstracts was significantly less than that of the non-relevant abstracts for eight out of the ten proteins (Mann-Whitney U test). 
However, the $p$-values may be underestimated because of the difference in the number of relevant and non-relevant abstracts for each protein. We observed a lower $p-$ value for proteins with fewer relevant abstracts.
Despite the high overlap between the two distributions we selected the open-weight Qwen3-Embedding:8B LLM as the embedder in the RAG framework for further analysis because it had the highest mean F1-score.


\subsection{Viral mutation extraction using RAG with full text} \label{results-rag-full-text}
We configured another RAG framework using the full textual information in the publications~(\sctn{\ref{methods:rag-full-text}}).
We divided the text in each publication into overlapping chunks.
Unlike RAG with abstracts, the datastore in this method contained embeddings of $10,327$ chunks of full text from $239$ influenza A publications.

\subsubsection{Evaluation of the extracted mutations} \label{results-rag-full-text-mutations}
As described in \sctn{\ref{results-rag-abstracts-mutations}}, we computed the precision, recall, and F1-scores of the retrieved mutations with respect to the ground truth mutations in the context~(\suppfig{\ref{appndx_fig:results-rag-fulltext-mutations-wrt-context-all-proteins-metrics}}) and overall ground truth~(\suppfig{\ref{appndx_fig:results-rag-fulltext-mutations-all-proteins-metrics}}).
Qwen3-Next-80B-A3B-Instruct model had the highest mean F1-score of $0.41\pm0.08$~(\fig{\ref{fig:results-vanilla-rag-f1scores}}).
The precision values with respect to ground truth in context were almost equal to the overall precision; and the recall values with respect to context were higher than the corresponding overall values~(\suppfig{\ref{appndx_fig:results-vanilla-rag-all-precision-recall}}).
This pattern, similar to the observation in RAG with abstracts, demonstrated the absence of LLM hallucinations of correct mutations. 

\subsubsection{Evaluation of the context} \label{results-rag-full-text-context}
We evaluated list of publications represented in the context through individual chunks.
Similar to RAG with abstracts, Qwen3-Embedding:8B had the best performance with average F1-score of $0.37\pm0.19$, followed by Open AI's text-embedding-3-large (\$) model with $0.33\pm0.17$~(\fig{\ref{fig:results-vanilla-rag-f1scores}}C).
We chose the open source Qwen3-Embedding:8B LLM as the embedder for further analysis again.

\subsection{Viral mutation extraction using \ourframework} \label{results-mlrag}
In this novel two-step RAG~(\sctn{\ref{methods:multi-level-rag}}), we harnessed the benefits of higher recall of relevant publications in the context using RAG with abstracts, and higher recall of viral mutations using RAG with full text.
Since the first level of selecting publications using abstracts was identical to the retrieval process in RAG with abstracts~(\sctn{\ref{results-rag-abstracts-context}}), we directly evaluate and discuss the performance of \ourframework in viral mutation extraction.
Similar to the previous two RAG frameworks, we used the Qwen3-Embedding:8B LLM as the embedder for subsequent evaluation of LLMs as responders in \mlr.

\subsubsection{Evaluation of the extracted mutations} \label{results-mlr-mutations}
\begin{figure*}[ht]
    \centering
    \includegraphics[width=\textwidth]{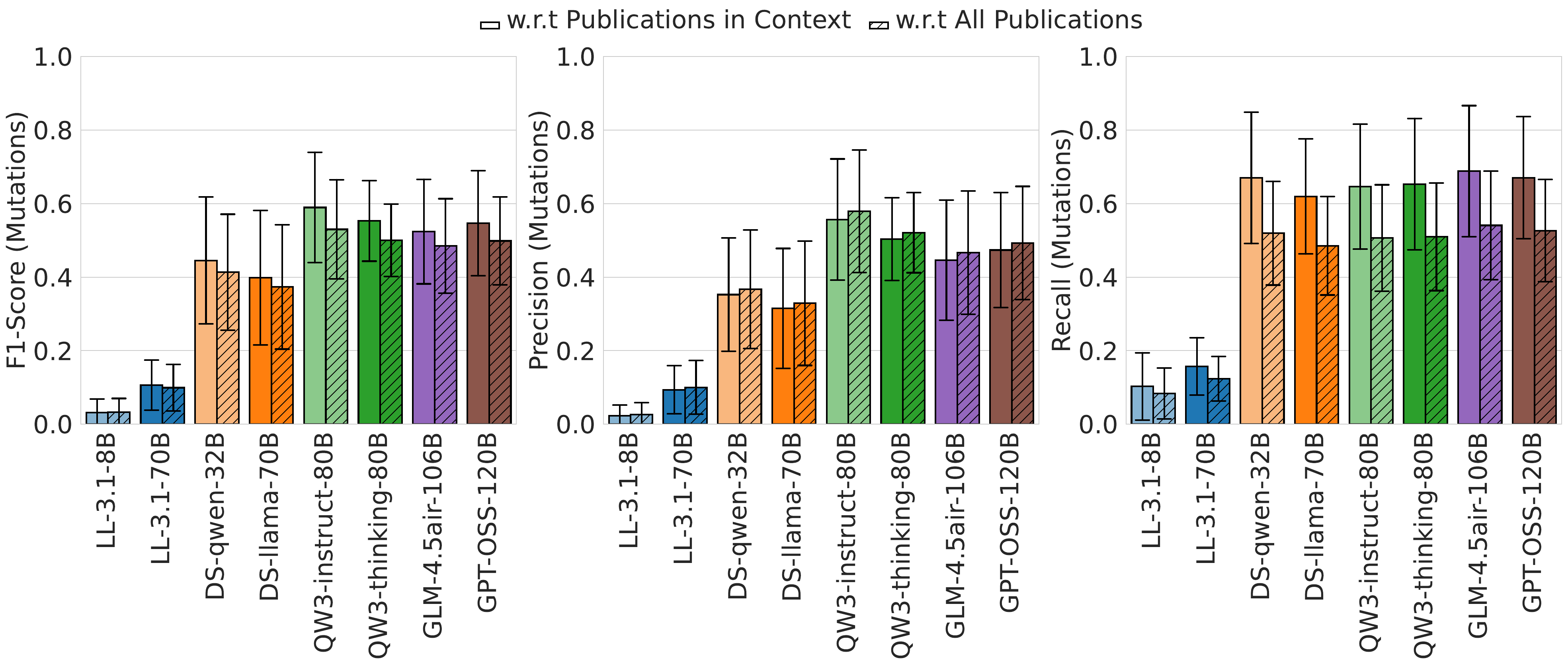}
    \caption[Evaluation of \ourframework for viral mutation extraction.]{Distribution of the precision, recall, and F1-scores with respect to the ground truth in the publications represented in the context in the prompt and the ground truth (hatched bars). The scores are for eight LLMs evaluated in identifying mutations across ten proteins in influenza A virus using \ourframework.}
    \Description{}
    \label{fig:mlr-paired-metrics}
\end{figure*}

We measured the precision, recall, and F1-scores of retrieved mutations with respect to the ground truth mutations corresponding to the provided context. We also computed the metrics with respect to all the ground truth mutations from all publications in datastore for a given protein~(\fig{\ref{fig:mlr-paired-metrics}}). 
Appendix figures~\ref{appndx_fig:results-multi-level-rag-mutations-all-proteins-metrics}~and~\ref{appndx_fig:results-multi-level-rag-mutations-wrt-context-all-proteins-metrics} contain the distribution of these metrics at the individual protein level.
\ourframework improved on the shortcoming of lower recall in the RAG with abstracts and full text methods.
Qwen3-Next-80B-A3B-Instruct had the highest mean F1-score of $0.53\pm0.13$ with similar mean precision ($0.58\pm0.17$) and recall ($0.50\pm0.14$) values.

\subsubsection{Selecting optimal number of abstracts and chunks in \ourframework}
We performed hyperparameter tuning to determine the optimal number of abstracts ($k_a$) and chunks from each publication ($k_c$) to be selected in the two retrieval steps of the \mlr algorithm~(Algorithm~\ref{alg:mlr}, \sctn{\ref{methods:multi-level-rag}}).
In this analysis, we used the LLMs with highest mean F1-score in selecting publications and retrieving mutations, as embedder and responder respectively. Thus, we used `Qwen3-Embedding-8B'~(\sctn{\ref{results-rag-abstracts-context}}) to compute embeddings and `Qwen3-Next-80B-A3B-Instruct' to generate the response~(\sctn{~\ref{results-mlr-mutations}}).
We used the influenza A dataset for viral mutation extraction in ten proteins with several combination of values for $k_a$ and $k_c$ (\suppfig{\ref{appndx_fig:results-mlr-hyperparam-k-metrics}}).
First, we explored six different values for $k_a$ from the set of $\{5, 10, 20, 40, 80, 160\}$ with $k_c$ fixed at $160$.
Next, we set $k_a=160$ and varied $k_c=\{5, 10, 20, 40, 80, 160\}$.

In \ourframework, $k_a$ controlled the number of publications selected in the first retrieval step~(\sctn{\ref{methods:multi-level-rag}}).
A larger value of $k_a$ enabled the framework to improve the recall but impacted the precision of the extracted mutations.
Starting at $k=5$ (precision$=0.44\pm40$), we observed highest mean precision of across all proteins at $k_a=20$ ($0.64\pm0.20$) which decreased by $\sim11\%$ at $k_a=160$ ($0.57\pm0.13$).
The recall steadily increased from $0.09\pm0.10$ to $0.50\pm0.14$ as we varied $k_a$ from $5$ to $160$.
Therefore there was an increasing trend in F1-scores as well from $0.14\pm0.15$ to $0.52\pm0.11$ with increasing $k_a$.

However, the values of $k_c$ did not influence the performance of \mlr method.
The mean F1-scores of retrieved mutations ranged from $0.52\pm0.12$ to $0.52\pm0.11$ when we varied $k_c$ from $5$ to $160$ .


\subsection{Benchmarking RAG-based frameworks for viral mutation extraction}
We compared the F1-scores of nine different LLMs as response generators (responders) for SIE using zero-shot prompting, two RAG-based methods, and \ourframework.
Our \mlr-based method significantly outperformed zero-shot prompting and RAG with abstracts for eight out of nine LLMs (\suppfig{\ref{appndx_fig:results-comparison-f1-score-all-villa-models}}).
Similarly, \ourframework was significantly better than RAG with full text for six LLMs.
Finally, we benchmarked state-of-the-at RAG-based tools for SIE namely, OpenScholar~\citep{asai-hajishirzi-openscholar-arxiv-2024}, PaperQA2~\citep{skarlinski-white-paperqa2-arxiv-2024}, and HiPerRAG~\citep{gokdemir-ramanathan-hiperrag-pascconf-2025}~(\fig{\ref{fig:teaser}}C, \tab{\ref{tab:villa-baseline-comparison}}, \suppfig{\ref{appndx_fig:sota-frameworks-metrics}}).
\begin{table*}[ht]
    \centering
    \begin{tabular}{|c|c|c|c|c|c|}
        \hline
        \textbf{SIE Method} & \textbf{Response Generator} & \textbf{Datastore ($\#$ papers)} & \textbf{F1-Score} & \textbf{Precision} & \textbf{Recall}\\
        \hline
        OpenScholar & Llama 3.1:8B & Semantic scholar (45 million) & $0.02\pm0.08$ & $0.04\pm0.07$ & $0.01\pm0.02$ \\
        \hline
        PaperQA2 (RAG mode) & Llama 3.1:8B & Influenza A ($239$) & $0.07\pm0.08$ & $0.27\pm0.30$ & $0.05\pm0.06$ \\ 
        PaperQA2 (Agent mode) & Llama 3.1:8B & Influenza A ($239$) & $0.05\pm0.06$ & $0.43\pm0.44$ & $0.03\pm0.04$ \\ 
        \hline
        HiPerRAG & Llama 3.1:8B & ASM journal ($\sim440$K) & $0.01\pm0.02$ & $0.04\pm0.07$ & $0.01\pm0.02$\\
        HiPerRAG & GPT-4o & ASM journal ($\sim440$K) & $0.09\pm0.12$ & $0.50\pm0.42$ & $0.05\pm0.07$\\
        HiPerRAG & GPT-4o & Influenza A ($239$) & $0.10\pm0.12$ & $0.52\pm0.39$ & $0.06\pm0.07$\\
        \hline
        VILLA (ours) & Llama 3.1:8B & Influenza A ($239$) & $0.03\pm0.04$ & $0.03\pm0.03$ & $0.08\pm0.07$\\
        VILLA (ours) & Qwen3-Instruct:80B & Influenza A ($239$) & \textbf{0.53$\pm$0.13} & \textbf{0.57$\pm$0.17} & \textbf{0.51$\pm$0.14}\\
        \hline
    \end{tabular}
    \caption{Mean and standard deviation of the distribution of the F1-scores, precision, and recall scores of OpenScholar, PaperQA2, HiPerRAG, and \ourframework frameworks in retrieving mutations of influenza A viral proteins. Each distribution is across five iterations of viral mutation extraction for each of the ten influenza A proteins.}
    \label{tab:villa-baseline-comparison}
\end{table*}

\subsection{Qualitative evaluation} 
\begin{figure}
    \centering
    \includegraphics[width=.9\linewidth]{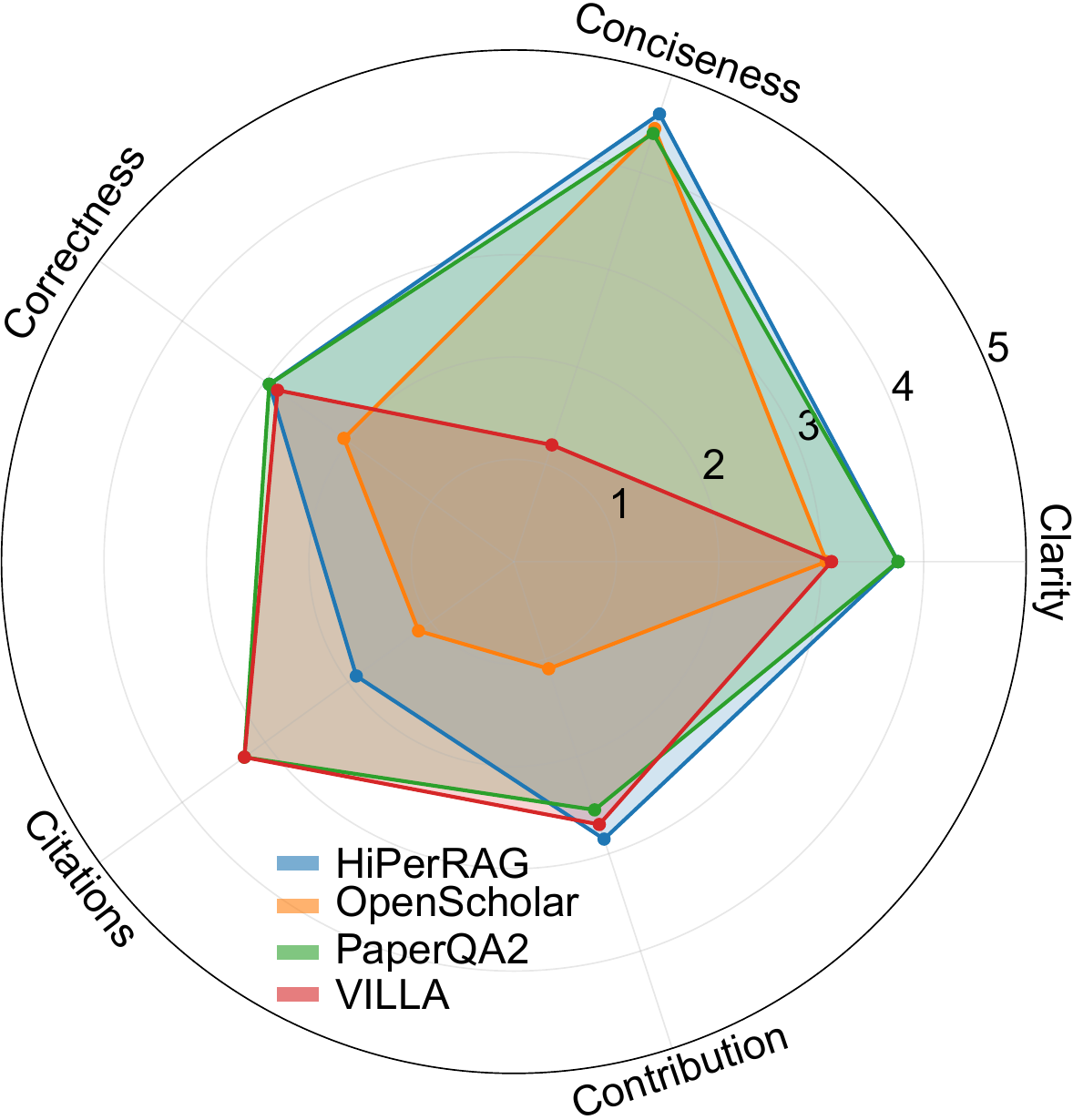}
    \caption[Qualitative evaluation of model responses]{Qualitative evaluation of the reasoning in model responses for the the viral mutation extraction task. Metric scores (5-point rubric) for each viral protein (\textit{n}=10) were coalesced into a single profile for each model framework.}
    \Description{}
    \label{fig:qualitative-evaluation}
\end{figure}

To assess the model outputs for their reasoning and insight, we devised a qualitative assessment to be performed by subject-matter experts (e.g., virologists). Our rubric utilized five metrics along two general dimensions: language/syntax (Correctness, Conciseness, Clarity) and biological significance (Citations and Contribution). See \suppnote{\ref{appndx_sec:qualitative_evaluation}} for details. We coalesced the respondent scores across each metric into a single profile for each model~(\fig{\ref{fig:qualitative-evaluation}}).
We observed that \ourframework diverged from the other models on the language/syntax dimension. Most models performed comparably while \ourframework lagged in Conciseness score. 
In contrast, \ourframework was comparable to other models along the dimension of biological relevance, with substantially improved performance of Citations.
\tmm{Can we briefly explain to the reader what Citations means here.}

Although we have established the substantial improvement of \ourframework in the quantitative measure of recall (\tab{\ref{tab:villa-baseline-comparison}}), our subject-matter experts observed limited improvement of \ourframework in `Contribution' (biological relevance) or `Correctness' (general language). We discuss the limitations of \ourframework highlighted by these trends in~Section~\ref{sec:limitations}. 
\section{Contribution to Virology}
\label{sec:domain-contribution}
Viruses are the most diverse and abundant biological entities with about $10^{31}$ viral particles in nature~\citep{paez-espino-kyrpides-virus-estimate-nature-2016, koonin-dolja-virus-estimate-environmentalmicrobiology-2023}. 
As many as $60\%$ of novel human infectious diseases since 1940 have been caused by viruses originating in animals~\citep{jones-daszak-infectious-diseases-trends-nature-2008, weiss-sankaran-zoonoses-emergence-f1000primerep-2022, forni-sironi-human-infecting-viruses-trendsinmicrobiology-2022}. 
Challenges in virology include foundational understanding of virus biology and pathogenesis, the development of therapeutic vaccines and anti-viral drugs, in addition to the development of predictive models of virus emergence and outbreak dynamics.
Computational methods, including ML models, have aided in the acceleration of research in virology~\citep{livesey-marsh-computational-variat-effect-predictor-molsystemsbiology-2020, beguir-sahin-earlywarningsystem-compbiomed-2023, rancati-marini-sarita-briefbioinform-2025, ito-sato-covfit-natcommunications-2025}.
There is a critical need for automated tools to accurately extract desired information from the vast scientific corpus and create structured datasets in virology that can drive research in AI for virology. 

To develop LLM-based SIE methods for virology, we used the influenza A virus to establish a controlled experiment of text retrieval on mutations in its proteins. We selected this virus since it has a rich literature over six decades.
In contrast, the research literature for a multitude of zoonotic- and pandemic-risk viruses are more dispersed and siloed. 
Unlike quantitative datasets in tabular and figure-borne formats, viral genome mutations (and their described impact on virus function) represent text-based information retrieval that is more nuanced. Therefore, any automated system of retrieval would be of significant value. Further, this data is critical to foundational questions of viral pathogenesis and emergence risk and may facilitate the development of new predictive models. 
Our results demonstrate that complex and multi-step methods are necessary to build a proper SIE protocol, thus tempering any over-confident rush to zero-shot methods. We also provide a framework for improved design and the need for iterative LLM pretraining data in domain-specific scientific literature.

\section{Limitations and Ethical Considerations}
\label{sec:limitations}

Since our research is on SIE, we do not see any implications related to consent or potential misuse. The data we use is present in the primary literature. We accessed the full text of relevant publications through institutional subscriptions to the relevant journals. 
\tmm{Should we say anything about licence issues for papers accessed through PaperQA?} We note that there may be an inherent bias in which proteins are studied by scientists~\citep{grudman-fiser-research-bias-iscience-2025}. In the case of influenza A virus, many more functionally-relevant mutations have been identified in the NA, PA, PB1, PB2 proteins.
\tmm{N, PA, PB1, PB2 proteins. Use full names. Refer to the relevant figure.} 
(\fig{\ref{fig:influenza-a-dataset}}).


Our quantitative results, F1-score of $0.53\pm0.13$, precision of $0.57\pm0.17$, and recall of $0.51\pm0.14$ (\tab{\ref{tab:villa-baseline-comparison}}), indicate that there is substantial ground for increasing the number of mutations retrieved correctly from the ground truth literature. Our analysis of the abstracts retrieved in the first step of \mlr demonstrated substantial overlap between the distribution of distances from the query to abstracts relevant to the corresponding protein and distribution of distances from the query to other abstracts~(\suppfig{\ref{appndx_fig:cosine-distance-abstracts-qwen3-8b}}). There is scope for considerable improvement in separating these distance distributions. 

Our qualitative results (\fig{\ref{fig:qualitative-evaluation}}) highlight additional limitations. The relatively poor evaluation of \ourframework in `Conciseness' results from its 
concatenation of the outputs from iterative queries for each relevant publication. We also observed limited improvement of \ourframework in `Contribution' (biological relevance) or `Correctness' (general language). This trend is likely a function of a limitation in our design where rubric-based qualitative evaluation of results can diverge from a quantitative score such as recall. Subject-matter experts do not inherently know the number of potential mutations of a protein present in the ground truth dataset. Therefore, an output of a small number of mutations (if accurate and accompanied by properly given context) cannot outscore a model result that correctly recalls even 3- or 4-fold greater mutations from the ground truth. While we can observe this disparity given our known ground truth dataset, SIE methods applied to topics without such a dataset may be challenged to demonstrate high recall in any limited qualitative assessment that may be desired. We hypothesize that this deficit, and a general tax from lack of conciseness, may have caused the parity of \ourframework with other models along the `biological relevance' dimension.

\section{Conclusion}
\label{sec:conclusion}
In this study, we defined a novel SIE task for viral mutation extraction.
We designed a novel \mlr method for SIE called \ourframework. 
Further, we curated a dataset of $629$ mutations across ten influenza A virus proteins from $239$ scientific publications.
We used this ground-truth dataset to evaluate a wide range of LLMs using baseline SIE methods such as zero-shot prompting and RAG- and agent-based tools.
We also demonstrated that \ourframework outperformed these baselines and had higher recall and F1-scores when compared to SOTA methods such as OpenScholar, PaperQA2, and HiperRAG.

We envision two future expansions of this study. First, a major challenge is to improve the performance (precision, recall, F1-score) of RAG-based approaches for our SIE task. One possibility may be to use a different prompt to select the right context. We may also consider NER methods to retrieve the correct chunks of textual information from publications.
Second, SIE methods typically involve querying the LLMs with one prompt at a time for a given task instance. In our case, the number of queries to the response generator LLM in the proposed \mlr approach was directly proportional to the number of abstracts selected in the first retrieval step. An important research question to address is systematically reducing the number of queries to LLMs in open-ended SIE tasks without compromising the quality of results.

\begin{acks}
National Science Foundation awards CCF-2412389 and CCF-2200045, the Pandemic Prediction and Prevention Destination Area at Virginia Tech, and the College of Engineering at Virginia Tech supported this research work.
\end{acks}
\bibliographystyle{ACM-Reference-Format}
\bibliography{references}

\appendix
\section{Related Work}\label{appndx_sec:relatedwork}
Initial works in SIE employed rule-based techniques, regular expressions, convolutional neural networks (CNN), and natural language processing methods such as recurrent neural networks (RNN) and long-short term memory (LSTM)~\citep{li-xu-sie-biomedical-jamia-2024, dagdelen-jain-sie-material-science-natcommunications-2024}.
SciFive~\citep{phan-altan-bonnet-scifive-arxiv-2021}, PubMedBERT~\citep{gu-poon-pubmedbert-blurb-acmtranscomputhealthcare-2021}, BioGPT~\citep{luo-liu-biogpt-briefbioinform-2022}, and MedCPT~\citep{jin-lu-medcpt-bioinformatics-2023} are some examples of LLMs that are pretrained on scientific literature in PubMed and used for scientific information extraction (SIE). These LLMs generate higher quality embeddings for biomedical text.

Common approaches for SIE include zero-shot prompting, few-shot prompting, retrieval augmented generation (RAG), and supervised fine-tuning (SFT).
In zero-shot prompting, we pose the task to the LLM in the form of a prompt in natural language. Here we do not retrain the LLM in any way or provide information in the form of labeled (or solved) examples. 
We rely on the ability of an LLM to steer its generated response based on guidance in the input prompt available via natural language.
Few-shot prompting involves adding one or more solved examples of the task within the prompt with the intention of guiding the LLM on how to respond.
In the RAG framework, an external tool is used to select pieces of text that are relevant to the query and may contain the potential answer from an external knowledge store. This retrieved `context' is used to augment the original prompt. The LLMs are expected to respond with the correct answer retrieved from the context.
SFT includes leveraging a ground truth dataset of question-answer pairs related to the task to fine-tune the parameters of LLM.
SFT updates the model weights through backward propagation of the loss incurred while identifying the correct answers to the queries in the ground truth dataset.

\section{Benchmark Datasets for Scientific Information Extraction} \label{appndx_sec:benchmark-datasets}
There are several standard datasets that are commonly used to benchmark LLMs in their ability to extract information from scientific publications. These datasets are set up for different types of scientific information extraction (SIE) tasks such as named entity recognition (NER), information extraction (IR), relation extraction (RE), document classification (DC), and question answering (QA). We describe in brief the different datasets used to evaluate LLMs in each of these tasks.
We also mention the proportion of `virus-related' records in these datasets from the ~\citet{chen-xu-benchmark-llm-biomedical-natcommun-2025} benchmarking study that included all standard datasets.
We tagged records containing any of the following terms as a virus-related record: `virus', `viruses', `viral', `virulent', `virulence'.

\subsection{Named entity recognition}
\paragraph{BC5-Chemical~\citep{li-lu-biocreative-vcdr-databaseoxford-2016, crichton-korhonen-bc5chem-bc5disease-datasets-bmcbioinformatics-2017, gu-poon-pubmedbert-blurb-acmtranscomputhealthcare-2021}.} Based on the BioCreative \romannumeraluppercase{5} Chemical-Disease Relation corpus of $1,500$ PubMed abstracts with annotations. The dataset contained $13,938$
records with chemicals (drugs) as named entities in these abstracts. There were $95$~(proportion$=0.68\%$) virus-related records in the dataset.

\paragraph{BC5-Disease~\citep{li-lu-biocreative-vcdr-databaseoxford-2016, crichton-korhonen-bc5chem-bc5disease-datasets-bmcbioinformatics-2017, gu-poon-pubmedbert-blurb-acmtranscomputhealthcare-2021}.} Based on the BioCreative \romannumeraluppercase{5} Chemical-Disease Relation corpus to create a dataset of $1,500$ PubMed abstracts with annotations. The dataset contained $13,938$ 
records with diseases as named entities in these abstracts. There were $95$~(proportion$=0.68\%$) virus-related records in the dataset.

\paragraph{NCBI-Disease~\citep{dogan-lu-ncbi-disease-dataset-biomedicalinformatics-2014, gu-poon-pubmedbert-blurb-acmtranscomputhealthcare-2021}.} Based on the Natural Center for Biotechnology Information Disease corpus containing $793$ PubMed abstracts with annotations. The dataset contained $7,287$ 
records annotated with $790$ unique disease entities. There were $25$~(proportion$=0.34\%$) virus-related records in the dataset.

\paragraph{BC2GM~\cite{smith-wilbur-bc2gm-dataset-genomebiology-2008, crichton-korhonen-bc5chem-bc5disease-datasets-bmcbioinformatics-2017, gu-poon-pubmedbert-blurb-acmtranscomputhealthcare-2021}.} Based on the Biocreative \romannumeraluppercase{2} Gene Mention corpus. The dataset contained $24,583$ records with manually annotated gene entities. Each record consisted of a sentence from PubMed abstracts and a set of gene entities mentioned in the sentence. 

\paragraph{JNLBPA~\citep{collier-kim-jnlbpa-dataset-nlpbabionlp-2004, crichton-korhonen-bc5chem-bc5disease-datasets-bmcbioinformatics-2017, gu-poon-pubmedbert-blurb-acmtranscomputhealthcare-2021}.} Based on the Natural Language Processing in Biomedicine and Its Applications Shared Task corpus. The dataset $59,963$ records of PubMed abstracts with labels as names of entities (such as cell line, cell type, DNA, protein, and RNA) in molecular biology.

\subsection{Information extraction}

\paragraph{EBM PICO~\citep{nye-wallace-ebmpico-dataset-acl-2018, gu-poon-pubmedbert-blurb-acmtranscomputhealthcare-2021}.} Based on the Evidence-Based Medicine corpus of $\sim 5,000$ abstracts on clinical trials in PubMed with annotations. There are $440,852$ records with abstracts and their corresponding annotations for the 
Participants, Intervention, Comparator, and Outcome.

\subsection{Relation extraction}

\paragraph{ChemProt~\citep{islamajdogan-lu-chemprot-dataset-databaseoxford-2019, gu-poon-pubmedbert-blurb-acmtranscomputhealthcare-2021}.} Based on the Chemical Protein Interaction corpus containing PubMed abstracts with annotations. The dataset had $48,223$ records where each record was a sentence from an abstract and a corresponding label with one of the following six interaction types between chemical and protein entities: `Upregulator', `Downregulator', `Agonist', `Antagonist', `Substrate', and `False' (for all other interactions). There were $207$~(proportion$=0.43\%$) virus-related records in the dataset.

\paragraph{DDI~\citep{herrero-zazo-declerck-ddi-dataset-biomedicalinformatics-2013, gu-poon-pubmedbert-blurb-acmtranscomputhealthcare-2021}.} Based on the Drug-Drug Interaction corpus containing PubMed abstracts with annotations. The dataset had $31,784$ 
records where each record was a sentence from an abstract. The boolean label in the record denoted whether the given sentence contained a drug-drug interaction. There were $241$~(proportion$=0.76\%$) virus-related records in the dataset.

\paragraph{GAD~\citep{bravo-furlong-gad-dataset-bmcbioinformatics-2015, gu-poon-pubmedbert-blurb-acmtranscomputhealthcare-2021}.} Based on the Genetic Association Database corpus containing sentences from PubMed abstracts with their corresponding gene-disease associations. The dataset has $5,330$ records where each record is a sentence from an abstract labeled with a gene-disease associations\todoblessy{add examples.}.

\subsection{Document classification}
\paragraph{HoC~\citep{hanahan-weinberg-hoc-dataset-cell-2011, gu-poon-pubmedbert-blurb-acmtranscomputhealthcare-2021}.} Based on the Hallmarks of Cancer corpus containing PubMed abstracts with their binary labels denoting the discussion of different cancer hallmarks. The dataset had $1,580$ 
records of PubMed abstracts annotated with binary labels for ten cancer hallmarks. There were $99$~(proportion$=6.27\%$) virus-related records in the dataset.

\paragraph{LitCovid~\citep{chen-lu-litcovid-dataset-nar-2021}.} The dataset contained abstracts of COVID-19 related publications from PubMed. Each article was labeled with one or more of the following eight categories using manual or computational methods: `General Information', `Mechanism', `Transmission', `Diagnosis', `Treatment', `Prevention', `Case Report', and `Epidemic Forecasting'. All the $33,699$ records in the dataset pertained to virology.

\subsection{Question answering}
\paragraph{PubMedQA~\citep{jin-lu-pubmedqa-dataset-emnlp-ijcnlp-2019}.} The dataset had $211,769$ 
records, each containing text from a PubMed abstract and a research question. The label (`Yes', `No', or `Maybe') in every record was the answer to the question based on the reference text. There were $11,275$ (proportion$=5.32\%$) virus-related records in the dataset.

\paragraph{BioASQ~\citep{nentidis-kakadiaris-bioasq-bionlpworkshopacl-2017}.} The dataset was adapted from the BioASQ corpus and had $885$ records. Each record contained text from a PubMed abstract and a research question. The label (`Yes' or `No') in every record was an answer to the question based on the reference text.

Existing studies benchmark LLMs using standard datasets spanning a wide range of topics such as drugs (BioCreative \romannumeraluppercase{5} Chemical-Disease Relation~\citep{crichton-korhonen-bc5chem-bc5disease-datasets-bmcbioinformatics-2017}, EBM PICO~\citep{nye-wallace-ebmpico-dataset-acl-2018}, ChemProt~\citep{islamajdogan-lu-chemprot-dataset-databaseoxford-2019}, DDI~\citep{herrero-zazo-declerck-ddi-dataset-biomedicalinformatics-2013}), diseases (NCBI-Disease~\citep{dogan-lu-ncbi-disease-dataset-biomedicalinformatics-2014}, HoC~\citep{hanahan-weinberg-hoc-dataset-cell-2011}, LitCovid~\citep{chen-lu-litcovid-dataset-nar-2021}), molecular biology (BC2GM~\citep{smith-wilbur-bc2gm-dataset-genomebiology-2008}, JNLBPA~\citep{collier-kim-jnlbpa-dataset-nlpbabionlp-2004}, GAD~\citep{bravo-furlong-gad-dataset-bmcbioinformatics-2015}), and general biomedical concepts (PubMedQA~\citep{jin-lu-pubmedqa-dataset-emnlp-ijcnlp-2019}, and BioASQ~\citep{nentidis-kakadiaris-bioasq-bionlpworkshopacl-2017}).
\section{Baseline Implementation} \label{appndx_sec:baselines}

\subsection{Zero-shot prompting}
We evaluated eleven (two closed and nine open) general purpose LLMs using zero-shot prompting for viral mutation extraction.

\subsection{RAG with abstracts}
An embedder LLM encoded these abstracts and stored the vector representations in a datastore.
We set the chunk size $s=5,000$ tokens (each token corresponds to a character) to avoid splitting a single abstract into two or more chunks.
Thus, $\mathcal{D_A}$ contained embeddings of $239$ abstract chunks.
At inference time, the embedder encoded the prompt.
We set $k_a=150$ and distance threshold$=0.5$ to retrieve the top $150$ abstracts closest and within a radius of $0.5$ to a given prompt in the embedding space.
Setting $k_a=150$ allowed the framework to capture information from a wide range of abstracts while adhering to the limitations on the size of the prompt enforced by the LLMs.
Further, we set the threshold $t=0.5$ to avoid selecting less relevant or non-relevant abstracts.
A retriever computed which abstracts had embeddings that were within a cosine distance of at most $0.5$ to that of the prompt's embedding.
The retriever chose a maximum of $150$ abstracts.
These selected abstracts constituted the context.
We augmented the original prompt with the concatenation of the selected abstracts.
A responder LLM received this augmented prompt containing instructions to extract the mutations for the given protein from the abstracts in the context.

\subsection{RAG with full text}
We first used the `marker' tool\footnote{https://github.com/datalab-to/marker} to parse the PDF files of the publications and extracted only the textual information from the main sections in the paper.
We did not include the abstract, references, or supplementary material.
We divided the full text of each publication into chunks of size $1,000$ tokens with an overlap of $100$ tokens.
Thus, $\mathcal{D_F}$ contained $10,327$ chunks of text from $239$ publications.
We set $k_c=150$ and distance threshold $=0.5$ to retrieve the top $150$ chunks closest and within a radius of $0.5$ to a given prompt in the embedding space.

\subsection{Response generator models} \label{appndx_sec:llms-generators}
\supptab{\ref{appndx_tab:generator-llms}} lists the different LLMs evaluated as response generators in zero-shot evaluation, RAG with abstracts, RAG with full text, and \ourframework SIE methods. 
\begin{table*}[ht]
    \centering
    \begin{tabular} {| l| l | l | l |}
    \hline
    Model Developer &  Model Name & Model Type & HuggingFace Model Name\\
    \hline
    \multirow{2}{0em}{DeepSeek-AI} & DeepSeek-R1-Distill-Llama-70B~\citep{deepseekai-deepseeekr1-arxiv-2025} & Reasoning & deepseek-ai/DeepSeek-R1-Distill-Llama-70B\\
    \cline{2-4}
     & DeepSeek-R1-Distill-Qwen-32B~\citep{deepseekai-deepseeekr1-arxiv-2025} & Reasoning & deepseek-ai/DeepSeek-R1-Distill-Qwen-32B\\
     \hline
     \multirow{2}{0em}{Meta} & Llama-3.1-8B & Instruction & meta-llama/Llama-3.1-8B\\
    \cline{2-4}
     & DeepSeek-R1-Distill-Qwen-32B & Reasoning & deepseek-ai/DeepSeek-R1-Distill-Qwen-32B\\
    \hline
    \multirow{3}{0em}{Qwen} & Qwen3-32B & Instruction & Qwen/Qwen3-32B\\
    \cline{2-4}
    & Qwen3-Next-80B-A3B-Instruct~\citep{qwen3-llms-arxiv-2025} & Instruction & Qwen/Qwen3-Next-80B-A3B-Instruct \\
    \cline{2-4}
    & Qwen3-Next-80B-A3B-Thinking~\citep{qwen3-llms-arxiv-2025} & Reasoning & Qwen/Qwen3-Next-80B-A3B-Thinking \\
    \hline
    \multirow{3}{0em}{OpenAI} & gpt-oss-120b~\citep{openai-zhao-openaigptoss120b-arxiv-2025} & Instruction & openai/gpt-oss-120b\\
    \cline{2-4}
     & GPT-4o (\$)~\cite{openai2024-openaigpt4o-2024} & Instruction & -\\
     \cline{2-4}
     & o1 (\$)~\citep{openai-li-openaio1-arxiv-2024} & Reasoning & - \\
    \hline
    \end{tabular}
    \caption{List of LLMs benchmarked as response generators in zero-shot prompting, RAG with abstracts, RAG with full-text, and \ourframework.}
    \label{appndx_tab:generator-llms}
\end{table*}

\subsection{Embedder models} \label{appndx_sec:llms-embedders}
\supptab{\ref{appndx_tab:embedder-llms}} contains the various embedding LLMs evaluated as embedders in RAG with abstracts, RAG with full text, and the proposed multi-level RAG frameworks.
\begin{table*}[ht]
    \centering
    \begin{tabular} {| l | r | r | l |}
    \hline
    Model Name & Context Length & Embedding Dimension & HuggingFace Model Name\\
    \hline
    Llama-3.1-8B & 128K & - & meta-llama/Llama-3.1-8B\\
    \hline
    \multirow{2}{0em}{MedCPT~(0.1B)~\citep{jin-lu-medcpt-bioinformatics-2023}} & 64 & 768 & ncbi/MedCPT-Query-Encoder\\
    \cline{2-4}
    & 512 & 768 & ncbi/MedCPT-Article-Encoder \\
    \hline
    Qwen3-embedding:8B~\citep{qwen3-llms-arxiv-2025} & 32K & 4,096 & Qwen/Qwen3-Embedding-8B\\
    \hline
    PubMedBERT~(0.1B)~\citep{gu-poon-pubmedbert-blurb-acmtranscomputhealthcare-2021} & & & NeuML/pubmedbert-base-embeddings \\
    \hline
    Open AI text-embedding-3-large & - & 3,072 & - \\
    \hline
    Google embeddinggemma-300M & 2,048 & 768 & google/embeddinggemma-300m \\
    \hline
    SFR-Embedding-Mistral~(7B) & 4,096 & 4,096\todoblessy{double check} & Salesforce/SFR-Embedding-Mistral\\
    \hline
    \end{tabular}
    \caption{List of LLMs benchmarked as embedders in RAG with abstracts, RAG with full-text, and \ourframework.}
    \label{appndx_tab:embedder-llms}
\end{table*}

\subsection{Other frameworks}
\paragraph{OpenScholar.} This enhanced RAG framework consists of a large-scale scientific datastore consisting of $45$ million open-access papers, dense retrievers further pre-trained to fetch relevant passages from datastore with scientific contents, cross-encoder re-ranker that was fine-tuned to select passages most relevant to the input query,
and response generation using customized and off-the-shelf LLMs. It also includes improvement of responses through iterative self-feedback triggering retrieval and inference.
We evaluated OpenScholar\footnote{https://openscilm.allen.ai/} through its user interface.

\paragraph{PaperQA2.} This model is a scientific question-answering framework that utilizes an agentic framework where an agent LLM acts as the `Tool Selector' and dynamically plans, selects tools, iteratively gathers and evaluates evidence, and decides when it has sufficient support to generate and terminate an answer. In its simple RAG mode, PaperQA2 executes a fixed, linear pipeline in which tools are invoked in a predefined order (search → evidence gathering → answer generation). 
\todoblessy{how did we use it?}

\paragraph{HiPerRAG}. This  modular, open-source high-performance computing (HPC) framework scales RAG across the full scientific pipeline, including PDF parsing, semantic chunking, embedding, retrieval, and answer generation.
The modular design of the framework enabled customization of each pipeline component, including AdaParse for PDF parsing~\citep{siebenschuh-underwood-adaparse-arxiv-2025}, PubMedBERT for semantic chunk boundary detection~\citep{gu-poon-pubmedbert-blurb-acmtranscomputhealthcare-2021}, SFR-Mistral-Embedding for chunk encoding~\citep{sfr-embedding-mistral-salesforce-2024}, and GPT-4o and LLaMA-3.1-8B as response generators.
In this study, we used HiPerRAG to construct a datastore of 448,650 scientific articles in microbiology curated in PDF format. 
The framework first parsed and extracted raw text using AdaParse~\citep{siebenschuh-underwood-adaparse-arxiv-2025}, a distributed PDF parsing pipeline that produces tokenizer-ready text at scale.

To address context-window constraints and restrict inference inputs to relevant content, we applied semantic chunking to the extracted corpus. This distributed chunking process ran on 8 NVIDIA A100 GPUs for approximately 18 hours, yielding 4,500,220 cogent and self-contained semantic text chunks. We used PubMedBERT \cite{gu-poon-pubmedbert-blurb-acmtranscomputhealthcare-2021} to encode sentence buffers during chunk boundary detection.
Next, we embedded each semantic chunk using SFR-Mistral-7B~\citep{sfr-embedding-mistral-salesforce-2024} to obtain $4096$-dimensional vector representations and indexed them in a FAISS vector store~\citep{johnson-jegou-faiss-big-data-ieeetxnsbigdata-2019} for efficient semantic search based on vector similarity.

\section{Qualitative Evaluation}\label{appndx_sec:qualitative_evaluation}
In the viral mutation extraction task, we prompted each model to retrieve the impact of each of the viral mutation on virus-host interaction apart from the mutations themselves.
The different models reported this information in field named `reasoning' which we manually evaluated to determine its quality and accuracy.
We defined five rubrics and developed a user interface for this qualitative evaluation process.

\subsection{Rubrics for evaluating LLM responses}
\todo{Add text here: Samantha, Brian}
To qualitatively assess the LLM reasoning responses, we developed a rubric with five categories: clarity, conciseness, correctness, citations/context, and contribution. 
All categories were measured on a five point scale, with one being the lowest score and five the highest.
The first category, clarity, addressed how clearly ideas and arguments are expressed in the reasoning, focusing on the organization and coherence of the response.
The second category, conciseness, addressed whether the reasoning text was efficient and purposeful to convey all relevant information.
While total length was a consideration, we did not want to penalize reasoning that included more mutations.
The third category, correctness, addressed whether the writing was mechanically correct and accurately identified mutations in the developed ground truth dataset.
The fourth category, citations/context, addressed whether the reasoning included citations and context that were relevant, authentic, and complete.
The fifth category, contribution, was a holistic measure of the usefulness of the reasoning and addressed whether the mutations and context provided in the reasoning were significant to the understanding of host-virus interactions and correctly answering the query.

\subsection{User interface for qualitative evaluation}
\label{appndx_sec:user-interface-qualitative-evaluation}
\begin{figure}[ht]
    \centering
    \includegraphics[width=\linewidth]{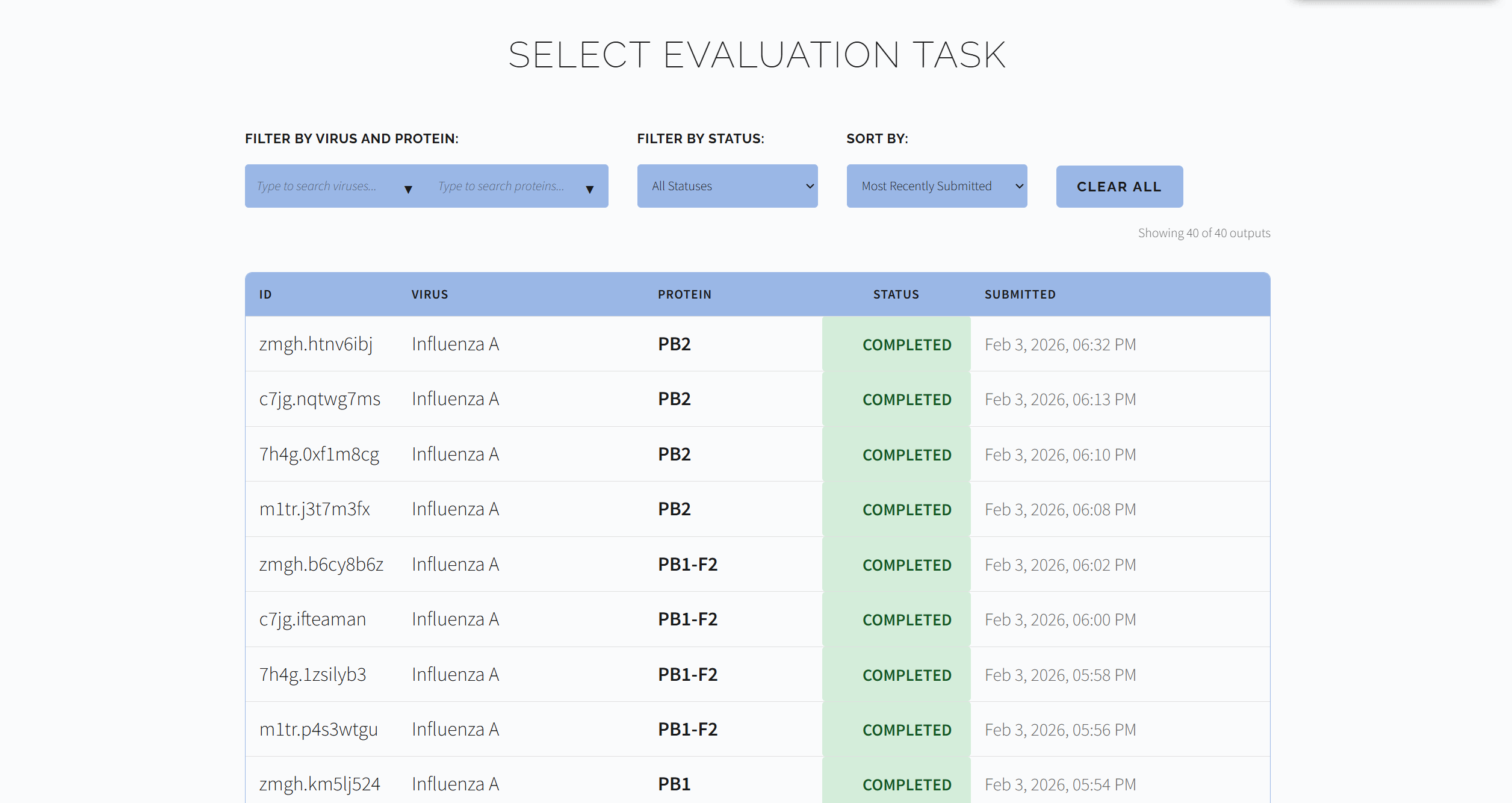}
    \caption{The overview page on evaluation interface allows evaluators to filter, sort, and select outputs for review while tracking evaluation progress and mitigate selection bias. Features include searchable virus/protein filters, status indicators (pending/completed), flexible sorting options, and anonymized output identifiers that prevent cherry-picking and ensure comprehensive evaluation coverage.}
    \Description{}
    \label{appndx_fig:ui_selection_page}
\end{figure}

\begin{figure}[ht]
    \centering
    \includegraphics[width=\linewidth]{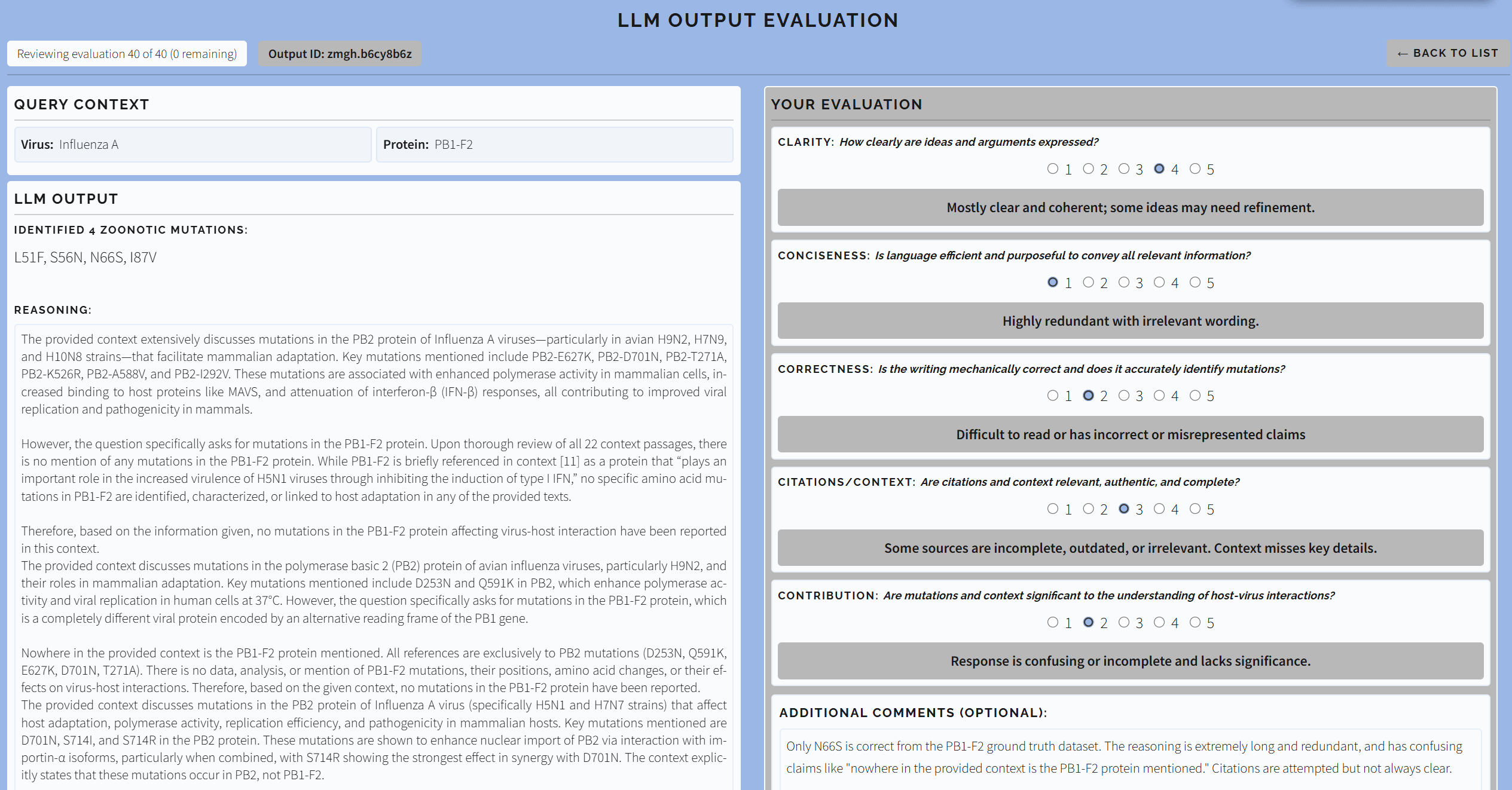}
    \caption{The evaluation interface for each LLM output consists of two primary components: a fixed left panel presenting extracted mutations and associated reasoning, and a right panel containing a structured Likert-scale rubric with dynamic descriptive text and an optional comment box for additional feedback.}
    \Description{}
    \label{appndx_fig:ui_eval_page}
\end{figure}

\begin{figure}[ht]
    \centering
    \includegraphics[width=\linewidth]{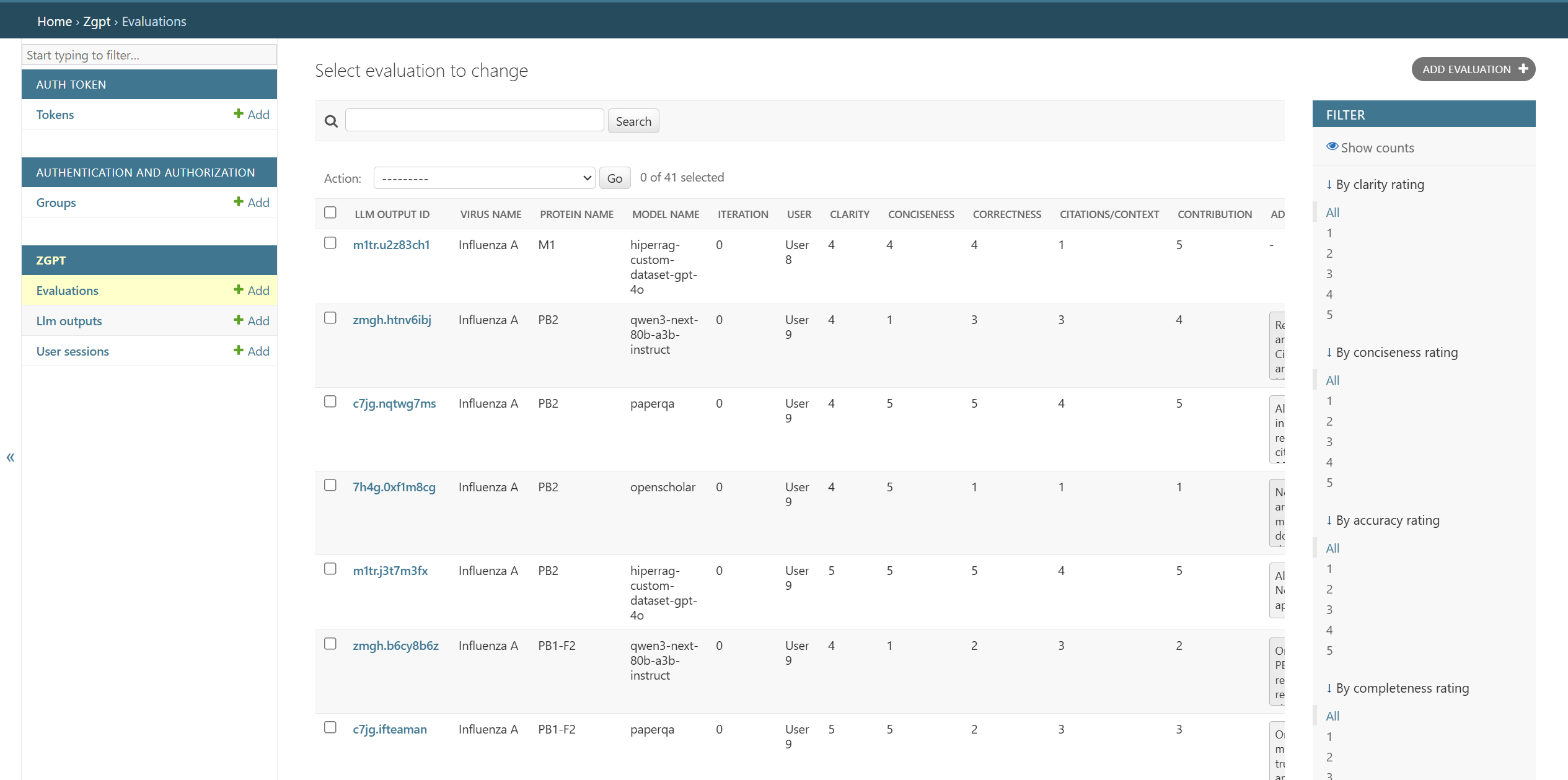}
    \caption{The admin dashboard enables research oversight with tabular views of LLM outputs and evaluations, filtering by criteria ratings and metadata, CSV export functionality for data analysis, and enabling monitoring for tracking evaluation progress and ensuring data quality throughout the study.}
    \Description{}
    \label{appndx_fig:ui_admin_page}
\end{figure}

We developed a web application to serve as the user interface for domain experts (virologists) to access, read, and evaluate the various model responses. \suppfig{\ref{appndx_fig:ui_selection_page}} and \suppfig{\ref{appndx_fig:ui_eval_page}} denote the interfaces to select a response of any model and evaluate the reasoning contents respectively. The administrator's interface track the evaluations of all domain experts is shown in \suppfig{\ref{appndx_fig:ui_admin_page}}.

\subsection{Implementation}
We used the open source persistent vector database called ChromaDb made available by LangChain\footnote{https://github.com/langchain-ai/langchain/tree/master/libs/partners/chroma} to construct the vector datastores in RAG with abstracts, RAG with full text, and \mlr experiments.
We used DSPy framework\footnote{https://github.com/stanfordnlp/dspy/tree/main} to implement the Zero-shot evaluation, RAG with abstracts, RAG with full-text, and \mlr methods.
Unless stated, all figures denote the distribution of metrics across five iterations of viral mutation extraction for each of the ten influenza A proteins.

\section{Viral Mutation Extraction Using Baseline Models}
We compared the performance of nine different LLMs as response generators (responders) for SIE using zero-shot prompting, two implementations of RAG, and \ourframework.
Additionally, we evaluated seven LLMs as embedders for the RAG-based methods designed in this study.

\subsection{Zero-shot prompting}
\label{appndx_sec:results-zero-shot-prompting}

\begin{figure}
    \centering
    \includegraphics[width=\linewidth]{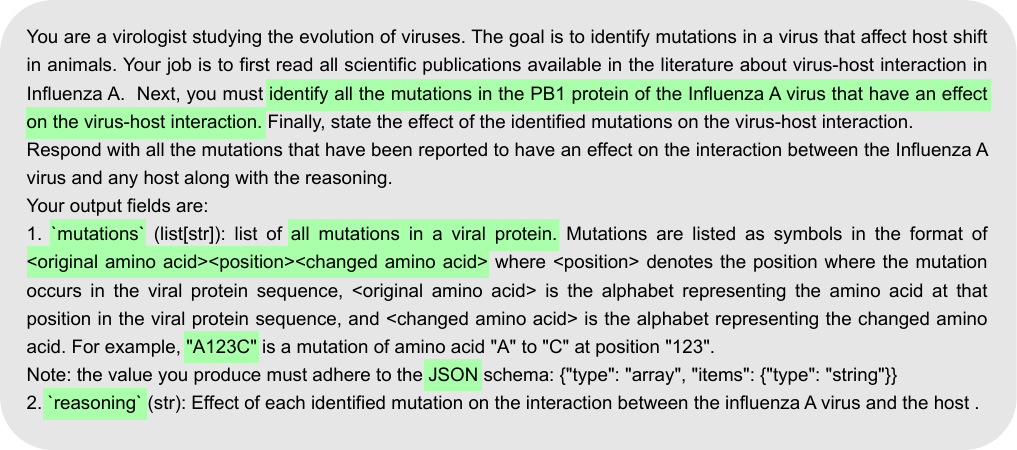}
    \caption{Prompt used to query large language models (LLMs) for viral mutation extraction using zero-shot prompting. The text highlighted in green denotes the key points such as description of the SIE task to respond with mutations impacting virus-host interaction in a given viral protein, the required output format, and the representation format of mutations.}
    \Description{}
    \label{appndx_fig:prompt-zero-shot}
\end{figure}

\begin{figure}[ht]
    \centering
    \includegraphics[width=\linewidth]{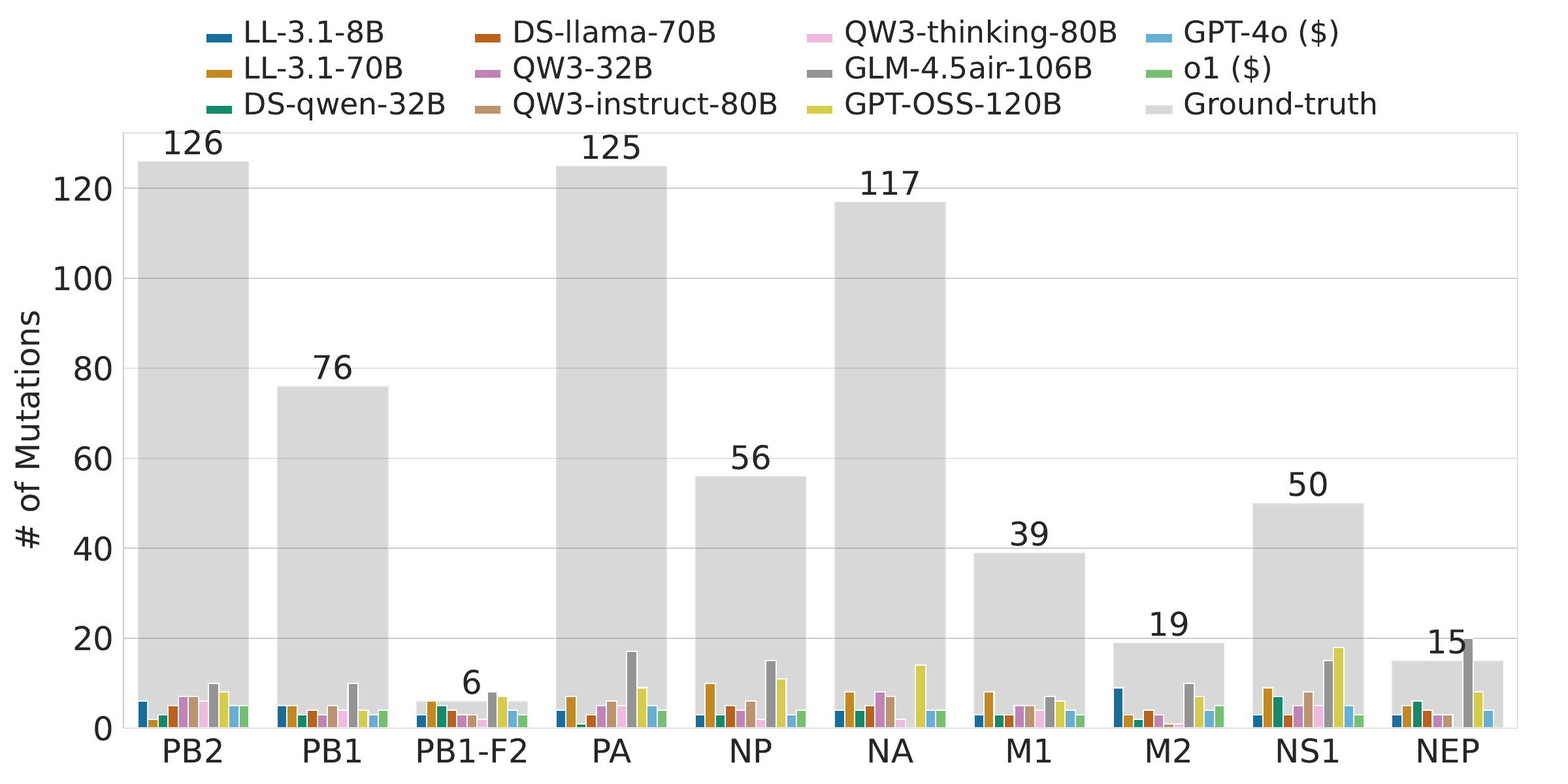}
    \caption{Distribution of the number of mutations identified by each of the eleven LLMs in ten proteins of the influenza A virus (\textit{x}-axis) using zero-shot prompting. The gray bars denote the number of mutations in the ground truth known to impact virus-host interactions. The \textit{x}-axis denotes the ten different influenza A proteins for which each LLM identified mutations. The height of each bar and the error mark in black correspond to the mean and standard deviation of the distribution respectively.}
    \Description{}
    \label{appndx_fig:results-zero-shot-groundtruth-retrieved-mutations-all-proteins}
\end{figure}

\begin{figure}[ht]
    \centering
    \includegraphics[width=\linewidth]{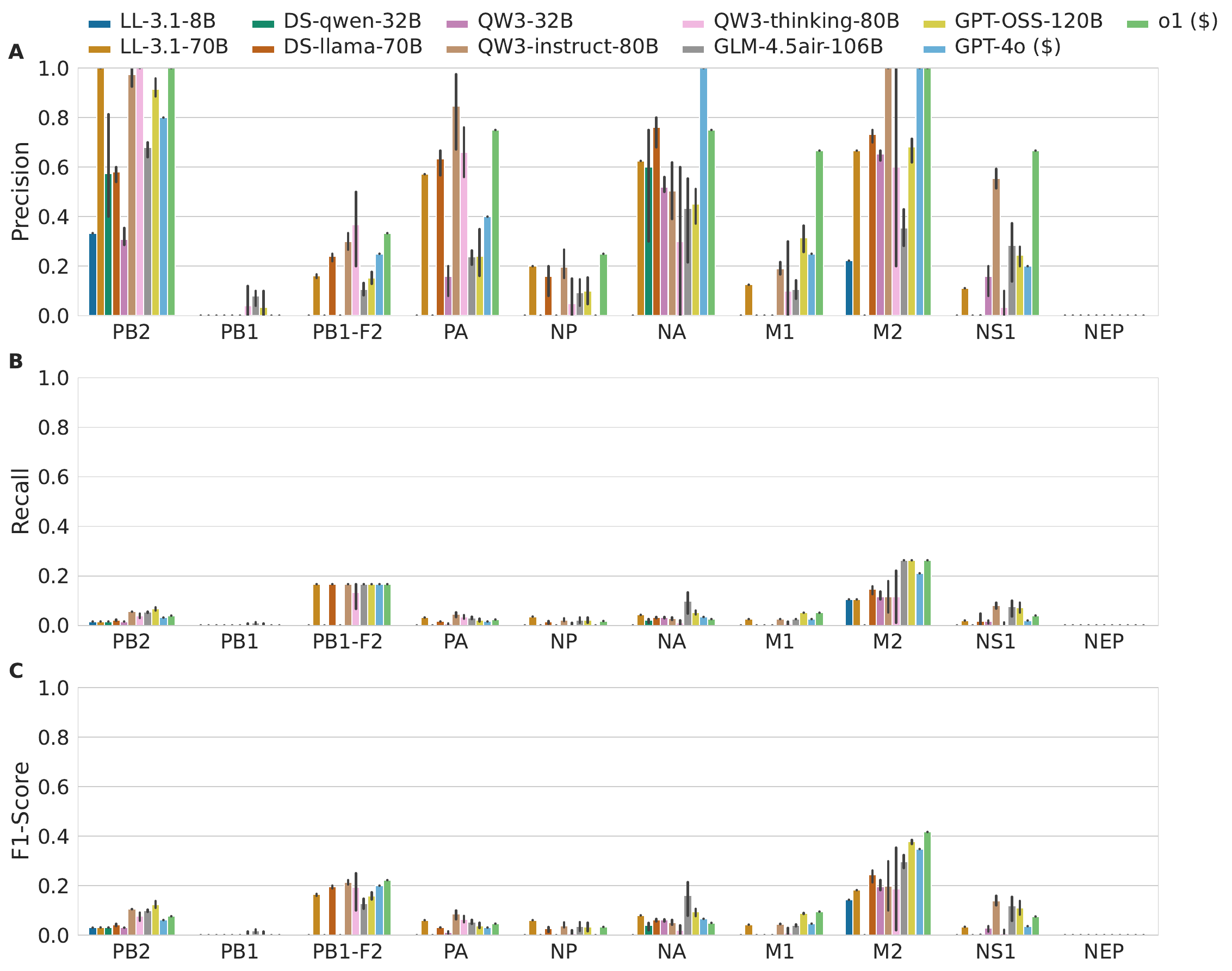}
    \caption{Distribution of the \textbf{(A)} precision, \textbf{(B)} recall, and \textbf{(C)} F1-scores of eleven general purpose LLMs in identifying mutations in influenza A virus that impact virus-host interaction. The \textit{x}-axis denotes the ten different influenza A proteins for which each LLM identified mutations. The height of each bar and the error mark in black correspond to the mean and standard deviation of the distribution respectively.}
    \Description{}
    \label{appndx_fig:results-zero-shot-evaluation-all-proteins-all-metrics}
\end{figure}

\suppfig{\ref{appndx_fig:prompt-zero-shot}} shows the prompt used to query responder LLMs in a zero-shot manner for viral mutation extraction.
\suppfig{\ref{appndx_fig:results-zero-shot-groundtruth-retrieved-mutations-all-proteins}} shows the distribution of the number of mutations identified by each LLM, as compared to ground truth mutations for each of the ten influenza A proteins.
The precision, recall, and F1-scores of LLMs under zero-shot evaluation varied substantially across all the proteins (\suppfig{\ref{appndx_fig:results-zero-shot-evaluation-all-proteins-all-metrics}}).

\subsection{Vanilla retrieval-augmented generation (RAG)}
\begin{figure}[ht]
    \centering
    \includegraphics[width=\linewidth]{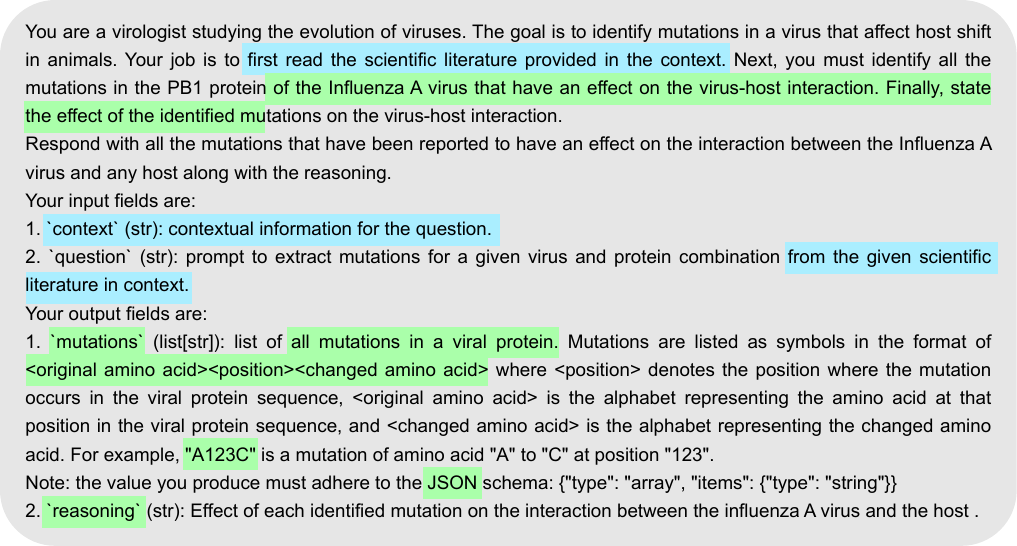}
    \caption{Prompt used to query large language models (LLMs) for viral mutation extraction using zero-shot prompting. The text highlighted in green denotes the key points such as description of the SIE task to respond with mutations impacting virus-host interaction in a given viral protein, the required output format, and the representation format of mutations. The text highlighted in blue instructs the LLMs to look only within the provided context while searching for answers.}
    \Description{}
    \label{appndx_fig:prompt-rag}
\end{figure}

We implemented two types of RAG with the datastore containing 
\begin{enumerate*}[(i)]
    \item embeddings of abstracts  and
    \item embeddings of chunks of full text from all 239 influenza A publications in ground truth dataset.
\end{enumerate*}
\suppfig{\ref{appndx_fig:prompt-rag}} shows the prompt in the RAG frameworks for viral mutation extraction. It includes instructions to extract the desired information from the provided context only.

\begin{figure}[ht]
    \centering
    \includegraphics[width=.5\textwidth]{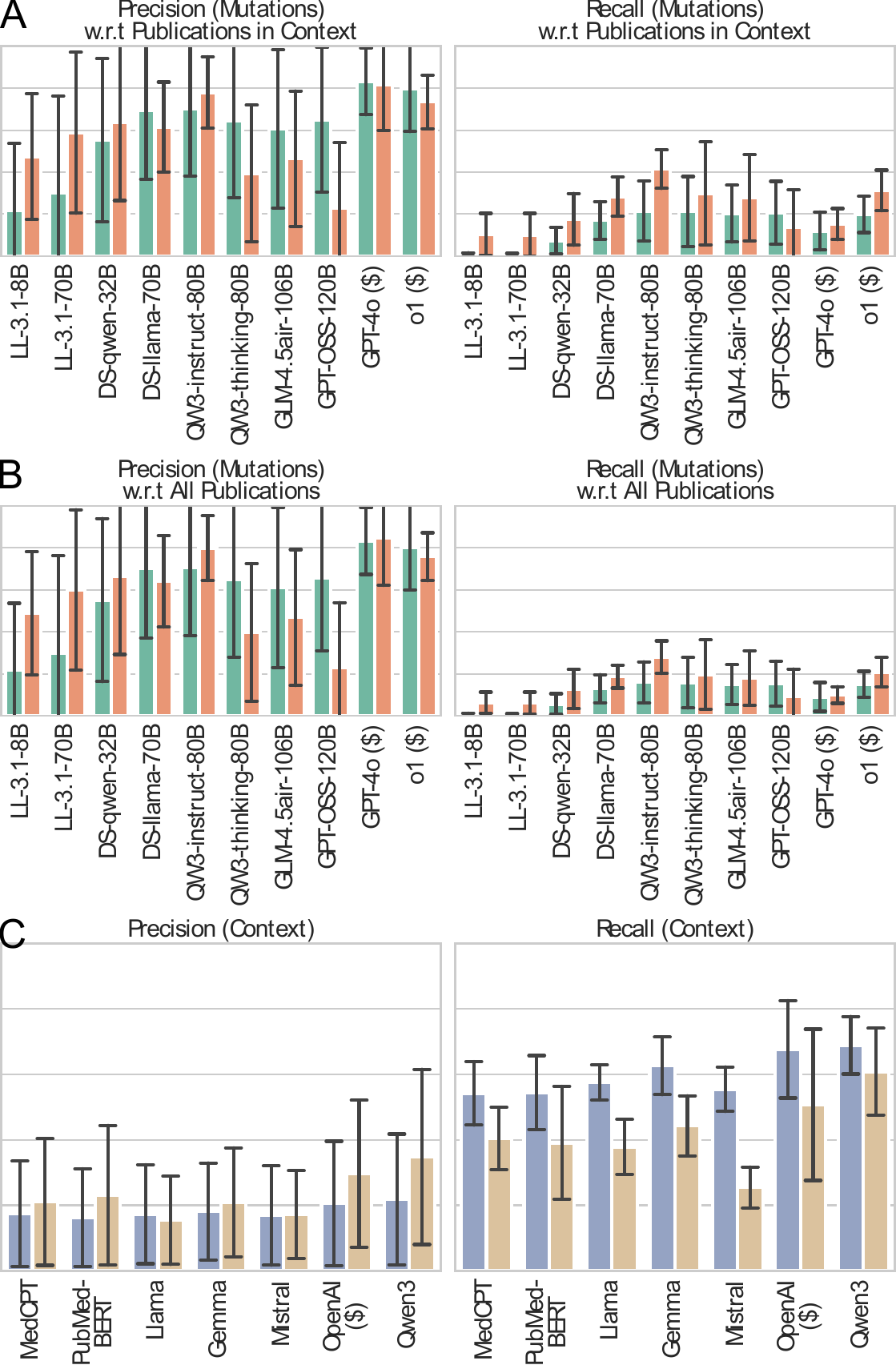}
    \caption[Evaluation of vanilla RAG frameworks for viral mutation extraction.]{Distribution of the \textbf{(A)} precision and recall with respect to the ground truth mutations in the publications represented in the context in the prompt and \textbf{(B)} the ground truth mutations in all the publications. The scores are for ten LLMs evaluated using RAG with abstracts and RAG with full text in identifying mutations across ten proteins in influenza A virus. 
    \textbf{(C)} Distribution of the precision and recall scores of the retriever in the two RAG frameworks in identifying abstracts relevant to each of the ten proteins in influenza A virus. The retriever is evaluated using seven different embedding LLMs.
    Each bar height and error bar corresponds to the mean and standard deviation of the distribution respectively. Abbreviations: LL, Llama; DS, DeepSeek; QW3, Qwen.}
    \Description{}
    \label{appndx_fig:results-vanilla-rag-all-precision-recall}
\end{figure}

\suppfig{\ref{appndx_fig:results-vanilla-rag-all-precision-recall}} denote the distribution of the precision and recall scores (averaged across ten influenza A proteins) of extracting the correct viral mutations using the two RAG implementations. It also contains the precision and recall scores of the retrievers for their ability to identify and include contents from the publications relevant to the given viral protein.

\subsubsection{RAG with abstracts}
\begin{figure}[ht]
    \centering
    \includegraphics[width=\linewidth]{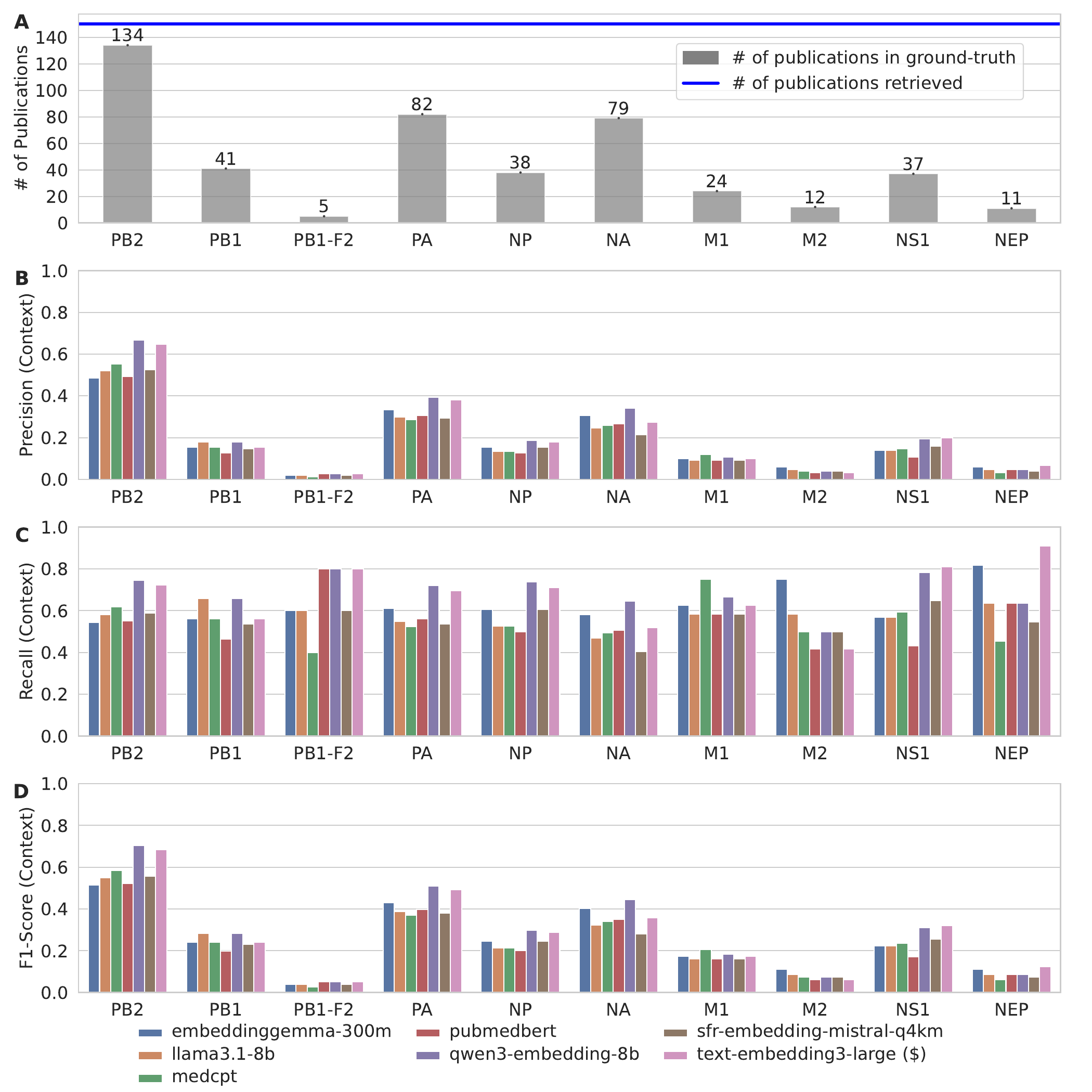}
    \caption{\textbf{(A)} The gray bars denote the number of abstracts in the ground truth relevant to each of the ten proteins (\textit{x}-axis). The solid blue line indicates the number of abstracts retrieved for each protein. Distribution of the \textbf{(B)} precision, \textbf{(C)} recall, and \textbf{(D)} F1-scores of the publications represented in the context of each prompt when querying to retrieve mutations in influenza A viral proteins using RAG with abstracts method. The performance of seven different embedding LLMs are compared through the evaluation of the context. The \textit{x}-axis denotes the ten different influenza A proteins for which each LLM identified mutations. The height of each bar and the error mark in black correspond to the mean and standard deviation of the distribution respectively.}
    \Description{}
    \label{appndx_fig:results-rag-abstracts-context-all-proteins-metrics}
\end{figure}

\begin{figure}[ht]
    \centering
    \includegraphics[width=\linewidth]{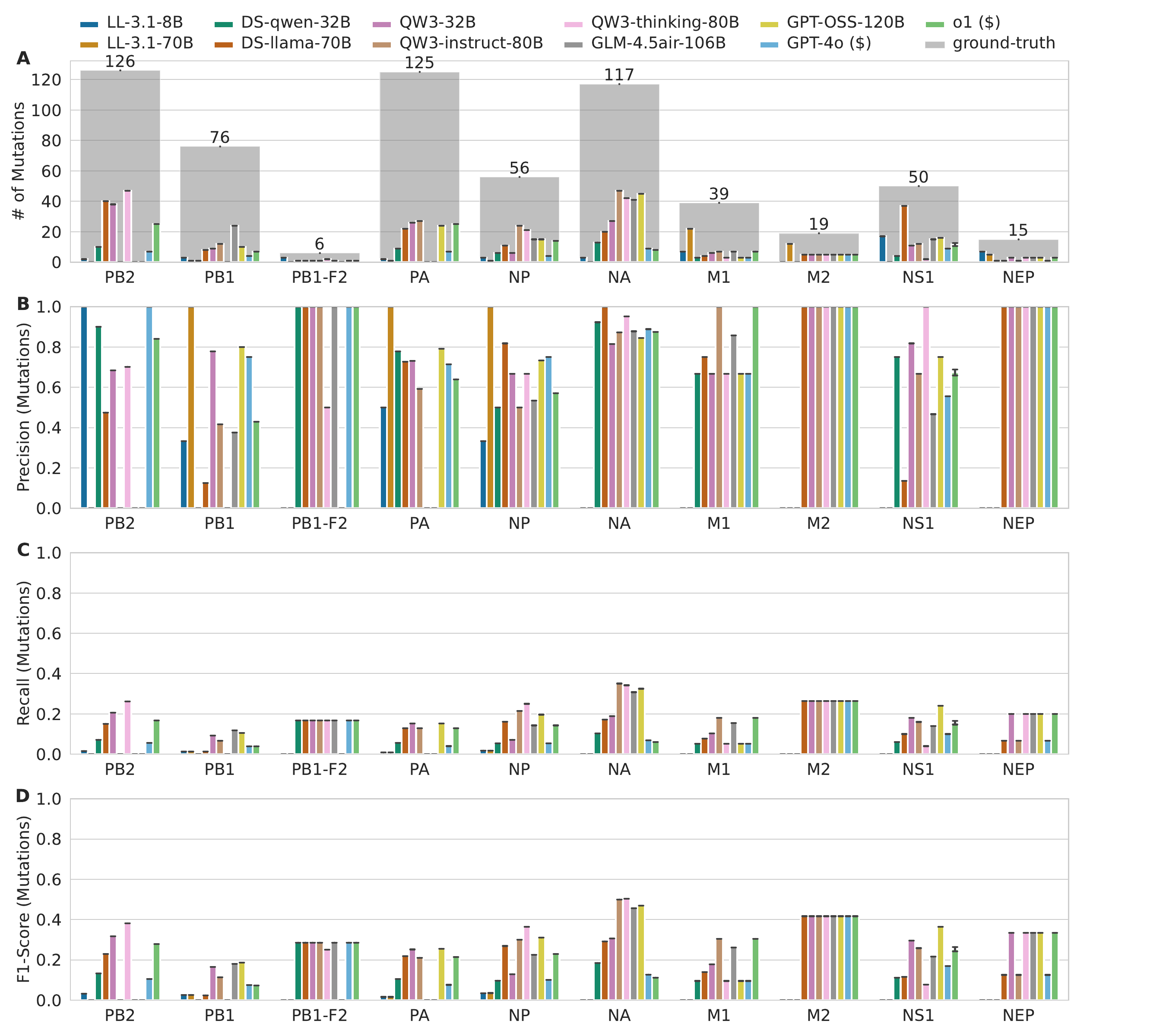}
    \caption{\textbf{(A)} Distribution of the number of mutations identified by each of the eleven LLMs in ten proteins of the influenza A virus (\textit{x}-axis) using RAG with abstracts. The gray bars denote the number of mutations in the ground truth known to impact virus-host interactions. Distribution of the \textbf{(B)} precision, \textbf{(C)} recall, and \textbf{(D)} F1-scores of eleven general purpose LLMs in identifying mutations in influenza A virus that impact virus-host interaction using RAG with abstracts method. The metrics are computed with respect to all the corresponding ground truth mutations of a given protein. The \textit{x}-axis denotes the ten different influenza A proteins for which each LLM identified mutations. The height of each bar and the error mark in black correspond to the mean and standard deviation of the distribution respectively.}
    \Description{}
    \label{appndx_fig:results-rag-abstracts-mutations-all-proteins-metrics}
\end{figure}

\begin{figure}[ht]
    \centering
    \includegraphics[width=\linewidth]{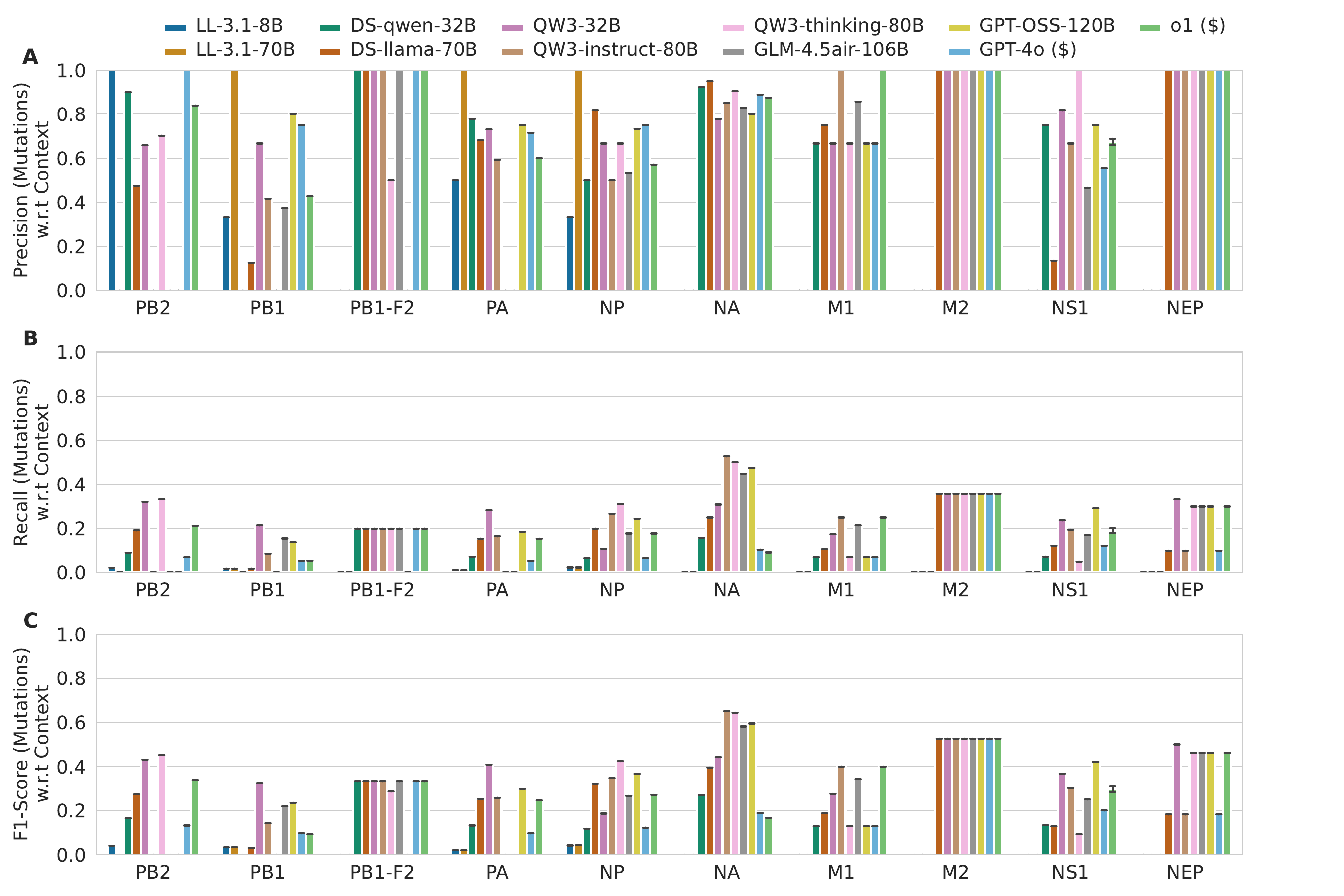}
    \caption{Distribution of the \textbf{(A)} precision, \textbf{(B)} recall, and \textbf{(C)} F1-scores of eleven general purpose LLMs in identifying mutations in influenza A virus that impact virus-host interaction using RAG with abstracts method. The metrics are computed with respect to the ground truth mutations in the publications represented in the context of each prompt. The distribution is across five iterations of SIE for each of the ten proteins. The \textit{x}-axis denotes the ten different influenza A proteins for which each LLM identified mutations. The height of each bar and the error mark in black correspond to the mean and standard deviation of the distribution respectively.}
    \Description{}
    \label{appndx_fig:results-rag-abstracts-mutations-wrt-context-all-proteins-metrics}
\end{figure}

This section illustrates the performance of the RAG framework using only the abstracts of the scientific publications in the dataset, including the number of abstracts retrieved per protein and the precision, recall and F1-scores of publications retrieved in the context (\suppfig{\ref{appndx_fig:results-rag-abstracts-context-all-proteins-metrics}}). We analyzed the number of mutations identified by different proteins for every protein and the corresponding precision, recall, and F1-scores for viral mutation extraction~(\suppfig{\ref{appndx_fig:results-rag-abstracts-mutations-all-proteins-metrics}}). We also evaluated the performance of mutation retrieval at per protein level with respect to the ground truth mutations in the publications represented in the context~(\suppfig{\ref{appndx_fig:results-rag-abstracts-mutations-wrt-context-all-proteins-metrics}}).

\subsubsection{RAG with full text}
\begin{figure}[ht]
    \centering
    \includegraphics[width=\linewidth]{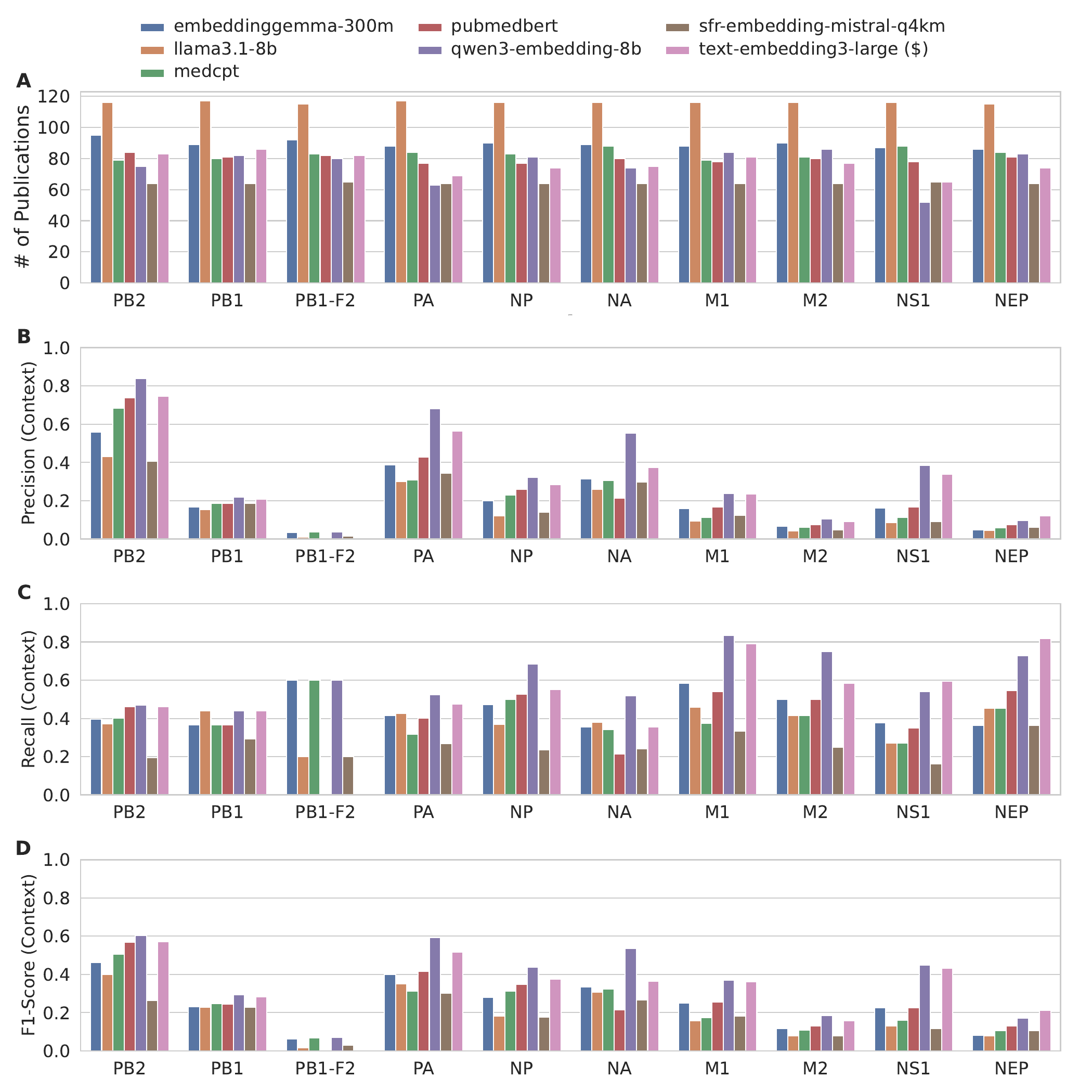}
    \caption{\textbf{(A)} Number of publications retrieved in the context for each of the ten influenza A proteins. The number in parentheses below every protein name in the \textit{x}-axis denotes the number of publications in the ground truth dataset relevant to each of the ten proteins. The colored bars show the number of publications identified by the retriever using the embeddings from seven different embedder LLMs. Distribution of \textbf{(A)} precision, \textbf{(B)} recall, and \textbf{(C)} F1-scores of the publications represented in the context of each prompt when querying to retrieve mutations in influenza A viral proteins using RAG with full text method. The performance of seven different embedding LLMs are compared through the evaluation of the context. The \textit{x}-axis denotes the ten different influenza A proteins for which each LLM identified mutations. The height of each bar and the error mark in black correspond to the mean and standard deviation of the distribution respectively.}
    \Description{}
    \label{appndx_fig:results-rag-fulltext-context-all-proteins-metrics}
\end{figure}

\begin{figure}[ht]
    \centering
    \includegraphics[width=\linewidth]{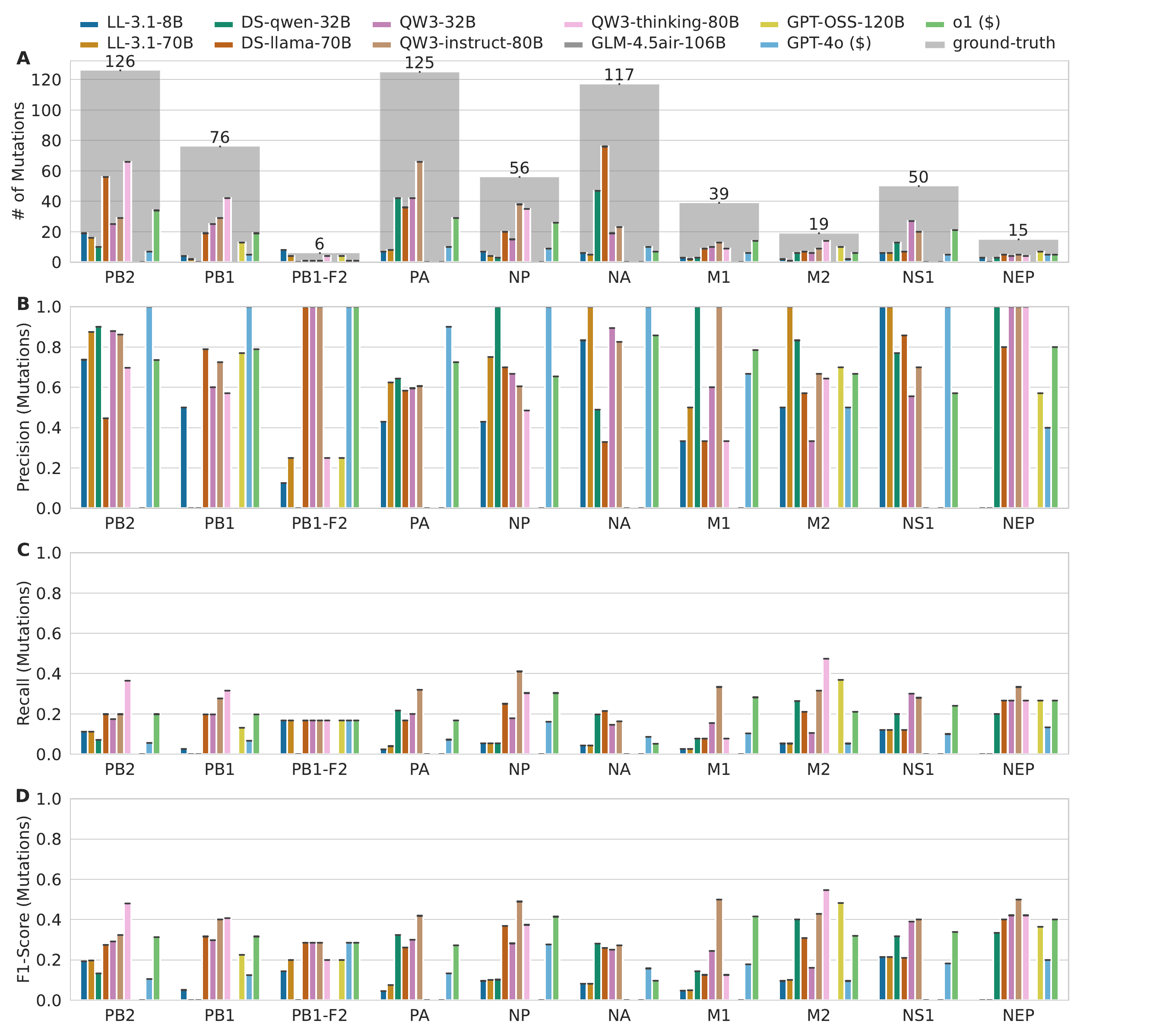}
    \caption{\textbf{(A)} Distribution of the number of mutations identified by each of the eleven LLMs in ten proteins of the influenza A virus (\textit{x}-axis) using RAG with full text of the publications. The gray bars denote the number of mutations in the ground truth known to impact virus-host interactions. Distribution of the \textbf{(B)} precision, \textbf{(C)} recall, and \textbf{(D)} F1-scores of eleven general purpose LLMs in identifying mutations in influenza A virus that impact virus-host interaction using RAG with full text method. The metrics are computed with respect to all the corresponding ground truth mutations of a given protein. The \textit{x}-axis denotes the ten different influenza A proteins for which each LLM identified mutations. The height of each bar and the error mark in black correspond to the mean and standard deviation of the distribution respectively.}
    \Description{}
    \label{appndx_fig:results-rag-fulltext-mutations-all-proteins-metrics}
\end{figure}

\begin{figure}[ht]
    \centering
    \includegraphics[width=\linewidth]{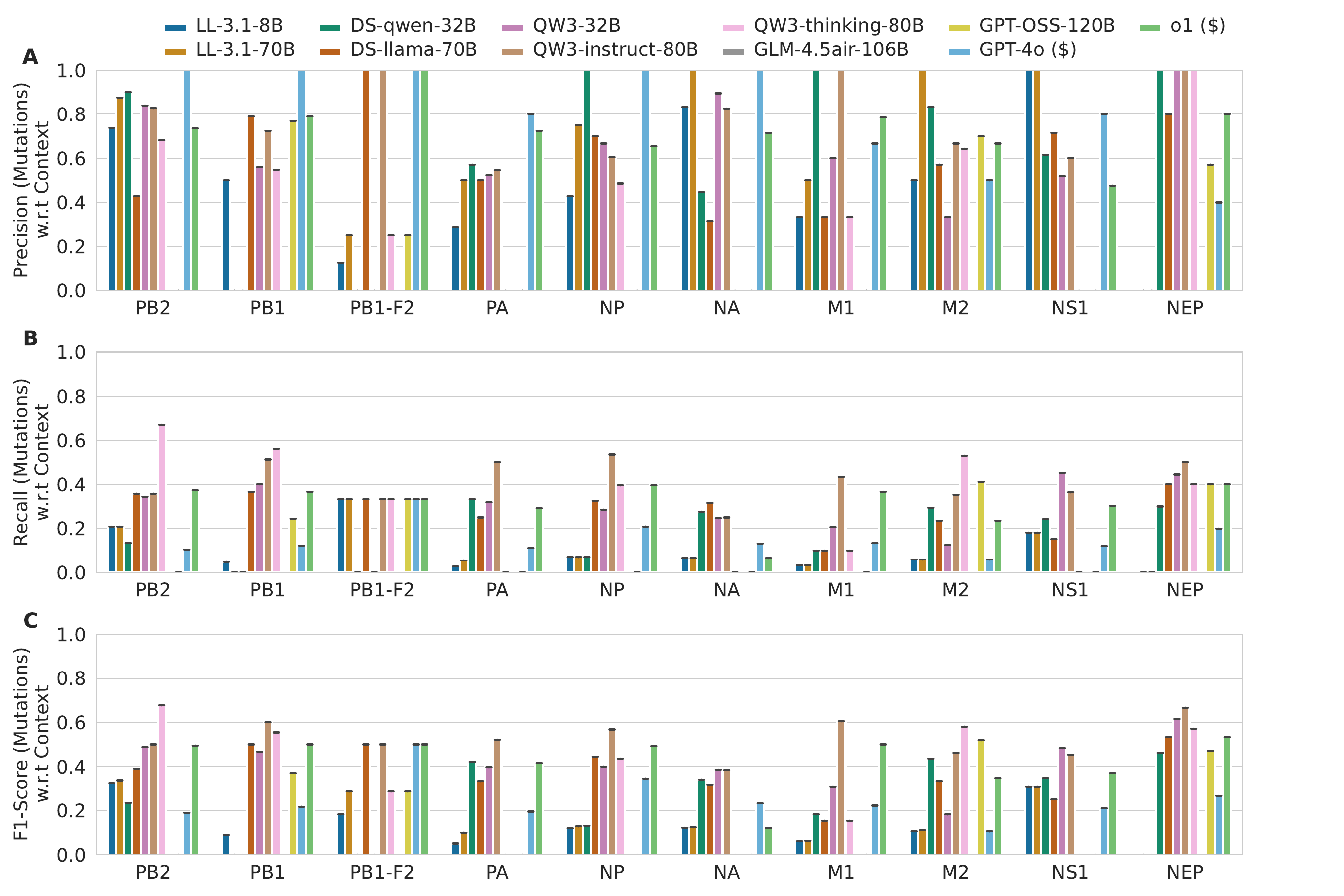}
    \caption{Distribution of the \textbf{(A)} precision, \textbf{(B)} recall, and \textbf{(C)} F1-scores of eleven general purpose LLMs in identifying mutations in influenza A virus that impact virus-host interaction using RAG with full text method. The metrics are computed with respect to the ground truth mutations in the publications represented in the context of each prompt. The \textit{x}-axis denotes the ten different influenza A proteins for which each LLM identified mutations. The height of each bar and the error mark in black correspond to the mean and standard deviation of the distribution respectively.}
    \Description{}
    \label{appndx_fig:results-rag-fulltext-mutations-wrt-context-all-proteins-metrics}
\end{figure}

This section illustrates the performance of the RAG framework using the full text of the scientific publications in the dataset, including the number of publications represented in the context through the retrieved chunks for each protein and the corresponding precision, recall and F1-scores~(\suppfig{\ref{appndx_fig:results-rag-fulltext-context-all-proteins-metrics}}). We analyzed the number of mutations identified by different proteins for every protein and the corresponding precision, recall, and F1-scores for viral mutation extraction~(\suppfig{\ref{appndx_fig:results-rag-fulltext-mutations-all-proteins-metrics}}). We also evaluated the performance of mutation retrieval at per protein level with respect to the ground truth mutations in the publications represented in the context~(\suppfig{\ref{appndx_fig:results-rag-fulltext-mutations-wrt-context-all-proteins-metrics}}).

\subsection{Viral mutation extraction using VILLA}

\begin{figure}[ht]
    \centering
    \includegraphics[width=\linewidth]{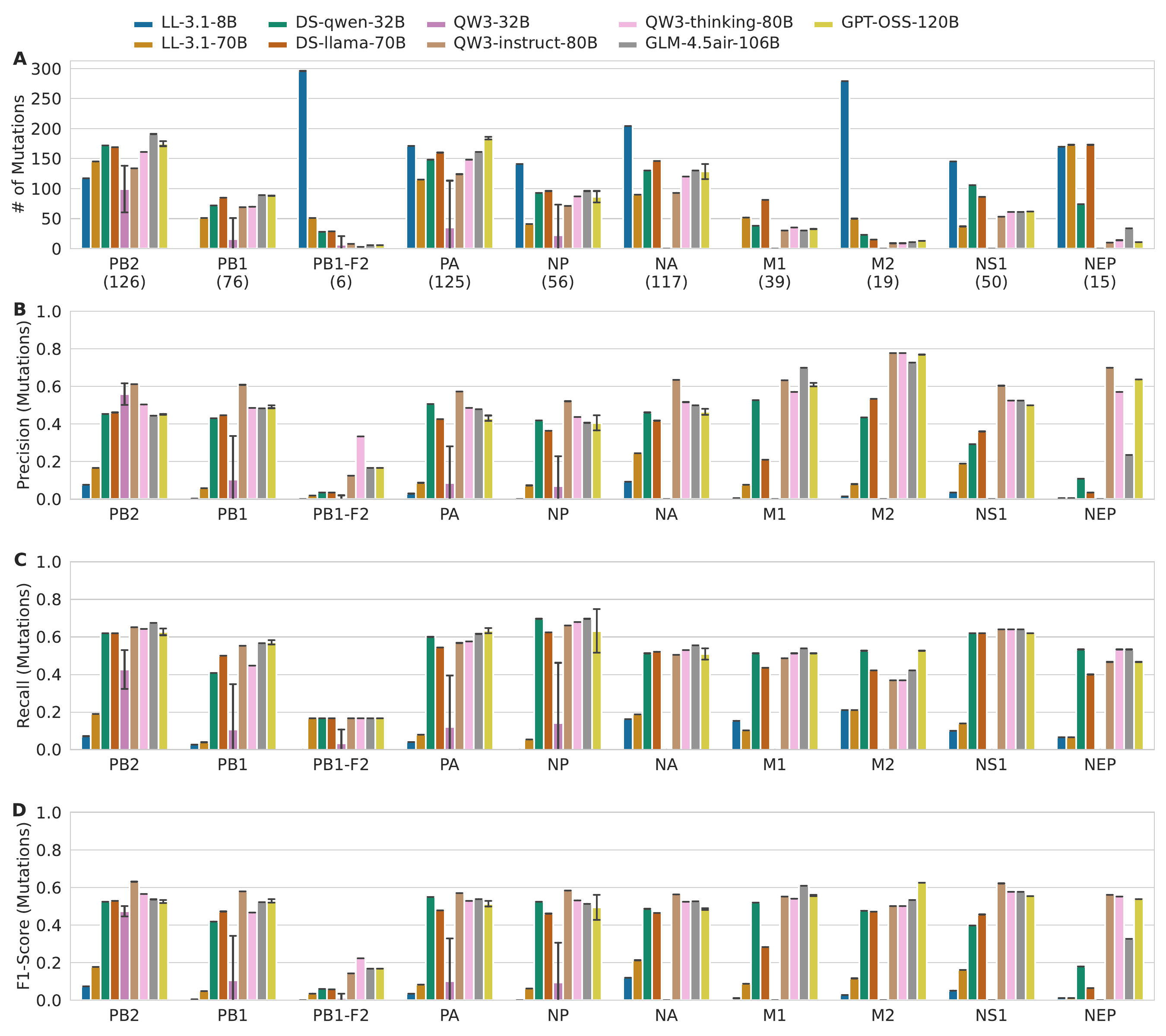}
    \caption{\textbf{(A)} Distribution of the number of mutations identified by each of the eleven LLMs in ten proteins of the influenza A virus (\textit{x}-axis) using RAG with full text of the publications. The number in parentheses below ever protein name in the \textit{x}-axis denotes the number of mutations in the ground truth known to impact virus-host interactions. Distribution of the \textbf{(B)} precision, \textbf{(C)} recall, and \textbf{(D)} F1-scores of eleven general purpose LLMs in identifying mutations in influenza A virus that impact virus-host interaction using \mlr method. The metrics are computed with respect to all the corresponding ground truth mutations of a given protein. The \textit{x}-axis denotes the ten different influenza A proteins for which each LLM identified mutations. The height of each bar and the error mark in black correspond to the mean and standard deviation of the distribution respectively.}
    \Description{}
    \label{appndx_fig:results-multi-level-rag-mutations-all-proteins-metrics}
\end{figure}

\begin{figure}[ht]
    \centering
    \includegraphics[width=\linewidth]{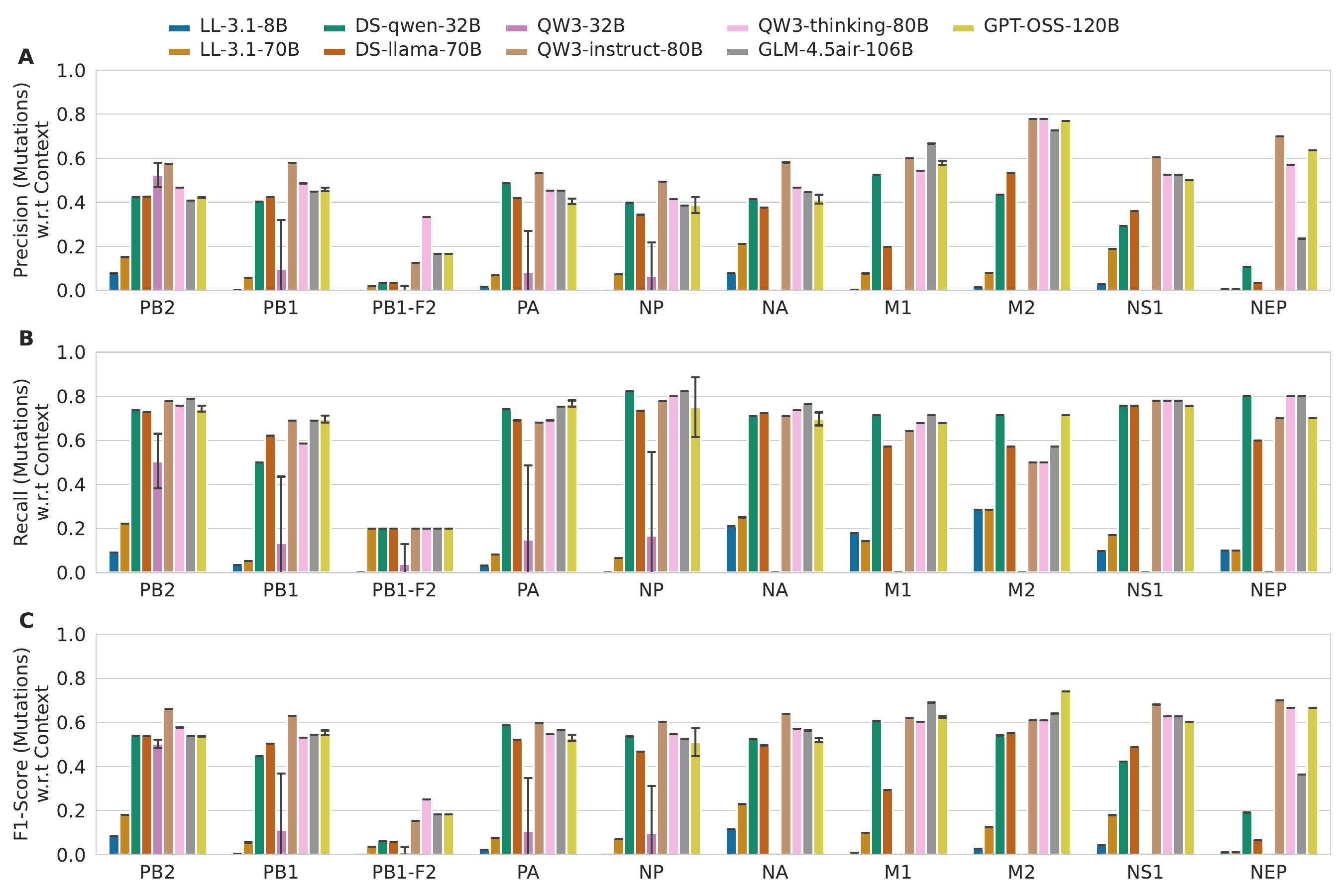}
    \caption{Distribution of the \textbf{(A)} precision, \textbf{(B)} recall, and \textbf{(C)} F1-scores of eleven general purpose LLMs in identifying mutations in influenza A virus that impact virus-host interaction using \mlr method. The metrics are computed with respect to the ground truth mutations in the publications selected in the first retrieval step. The \textit{x}-axis denotes the ten different influenza A proteins for which each LLM identified mutations. The height of each bar and the error mark in black correspond to the mean and standard deviation of the distribution respectively.}
    \Description{}
    \label{appndx_fig:results-multi-level-rag-mutations-wrt-context-all-proteins-metrics}
\end{figure}

This section illustrates the performance of the \mlr framework involving both abstracts and the full text of the scientific publications in the dataset.
We analyzed the number of mutations identified by different proteins for every protein and the corresponding precision, recall, and F1-scores for viral mutation extraction~(\suppfig{\ref{appndx_fig:results-multi-level-rag-mutations-all-proteins-metrics}}).
We also evaluated the performance of mutation retrieval at per protein level with respect to the ground truth mutations in the publications represented through the abstracts identified in the level one of retrieval~(\suppfig{\ref{appndx_fig:results-multi-level-rag-mutations-wrt-context-all-proteins-metrics}}).

\begin{figure}
    \centering
    \includegraphics[width=\linewidth]{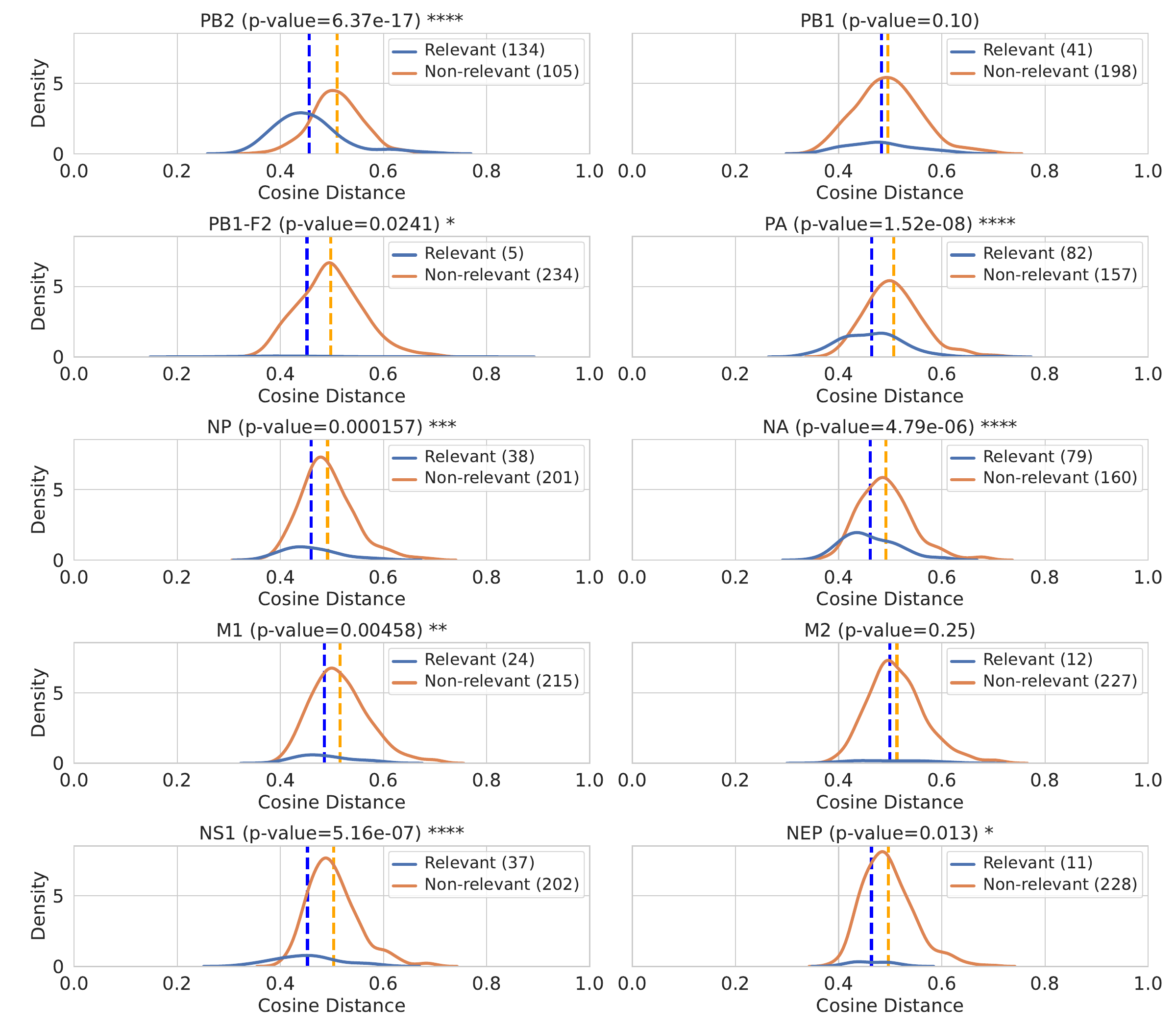}
    \caption[Analysis of distances of abstract embeddings from viral mutation extraction prompts.]{Comparison of the distribution of cosine distances between the embeddings of abstracts and the embedding of a prompt for viral mutation extraction in given influenza A protein. Each subplot corresponds to one of the ten proteins. The dashed vertical lines in blue and orange denote the mean cosine distance of the relevant and non-relevant abstracts from the prompt respectively. All embeddings were computed using Qwen3-Embedding:8B LLM. For each protein, the $p$-value denotes the statistical significance of the two distributions begin different.\todoblessy{update p-values in mathematical form}}
    \Description{}
    \label{appndx_fig:cosine-distance-abstracts-qwen3-8b}
\end{figure}

\begin{figure*}[ht]
    \centering
    \includegraphics[width=\linewidth]{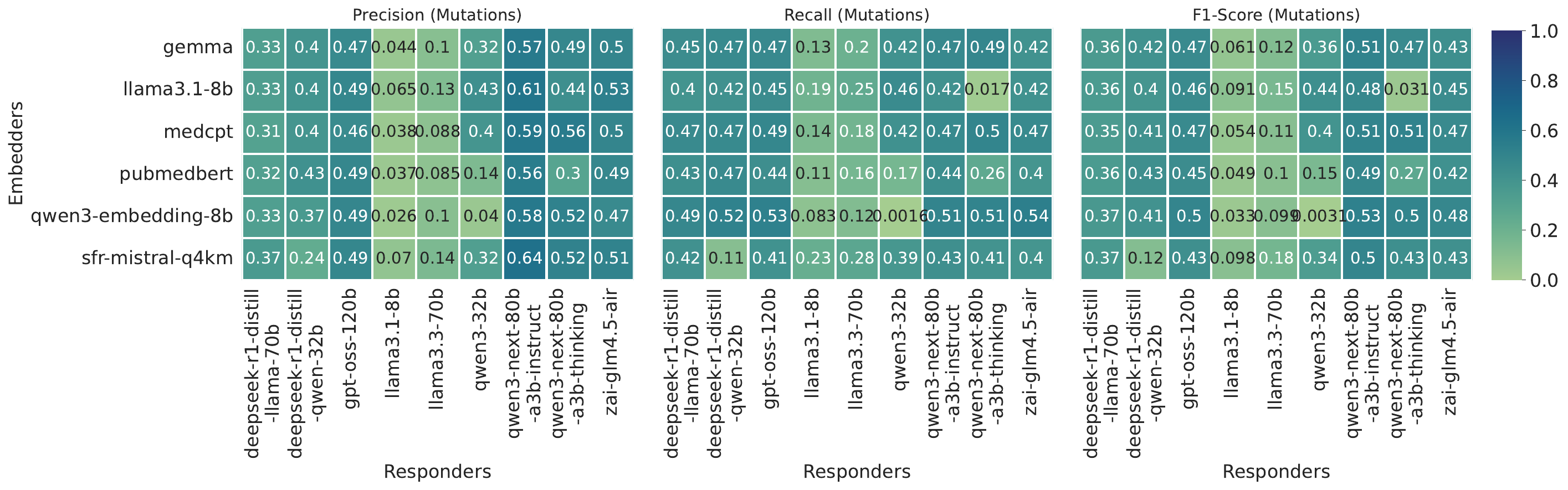}
    \caption{Performance of different combinations of LLMs as embedders and response generators in identifying mutations in influenza A virus that impact virus-host interaction using \mlr method. In each heatmap, the rows correspond to six different embedder models, and the columns denote nine different response generators. Each cell shows the average precision, recall, or F1 score across five iterations of SIE for each of the ten influenza A proteins.}
    \Description{}
    \label{appndx_fig:results-multi-level-rag-all-emb-gen-combinations}
\end{figure*}

\begin{figure}[ht]
    \centering
    \includegraphics[width=\linewidth]{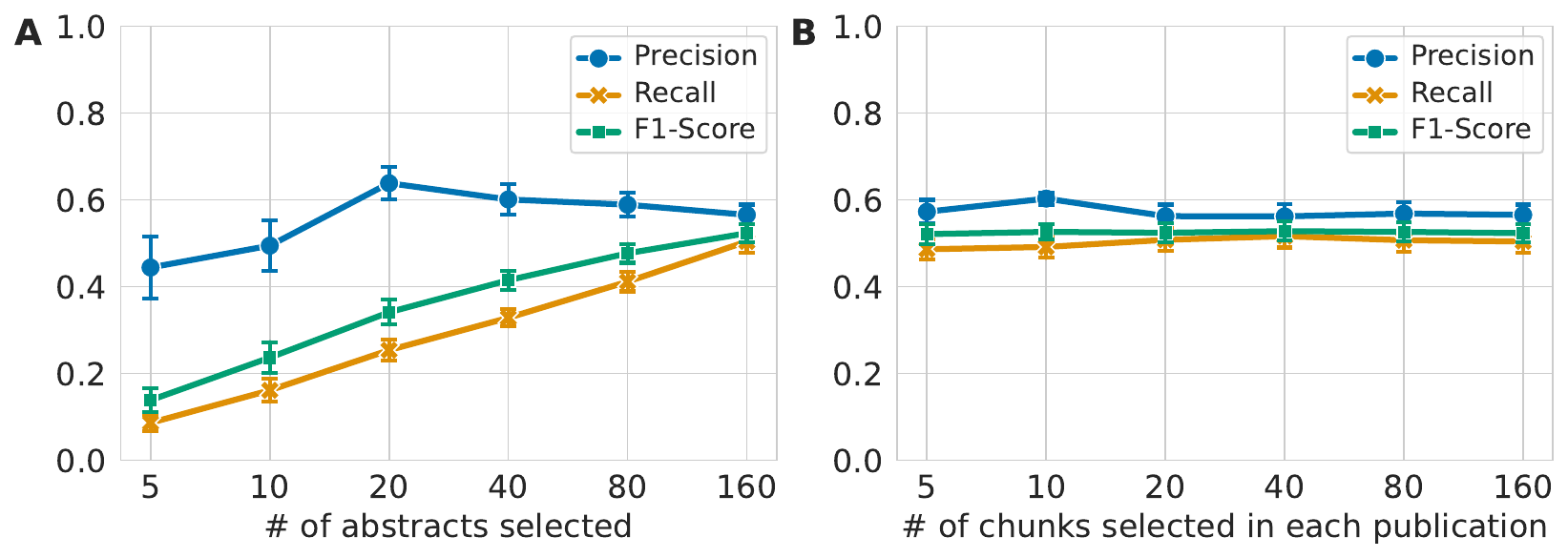}
    \caption{\textbf{(A)} Precision, recall, and F1-Scores for different values of number of abstracts selected in \mlr with the number of chunks selected in each publication fixed at $160$. \textbf{(B)} Precision, recall, and F1-Scores for different values of number of chunks selected from each publication in \mlr with the number of abstracts selected fixed at $160$. In both sets of experiments, the embedder and responder models were `Qwen3-Embedding-8B' and `Qwen3-Next-80B-A3B-Instruct' respectively. The scores are for identifying mutations across ten proteins in Influenza A virus. The \textit{x}-axis denotes the different values of the hyperparameter. Each point in the line and the error bar denote the mean and standard deviation of the corresponding distribution respectively.}
    \Description{}
    \label{appndx_fig:results-mlr-hyperparam-k-metrics}
\end{figure}

We used Qwen3-Embedding:8B as the embedder LLM to compare the distribution of cosine distances between the embeddings of abstracts and the embedding of a prompt for viral mutation extraction in each influenza A protein~(\suppfig{\ref{appndx_fig:cosine-distance-abstracts-qwen3-8b}}).
\suppfig{\ref{appndx_fig:results-multi-level-rag-all-emb-gen-combinations}} shows the performance of different combinations of six LLMs as embedders and nine LLMs as response generators in identifying mutations in influenza A virus that impact virus-host interaction using \mlr method. Additionally, we performed hyperparameter analysis to determine the optimal values for number abstracts ($k_a$), and number of chunks ($k_c$) in the two levels of retrieval in \ourframework respectively~(\suppfig{\ref{appndx_fig:results-mlr-hyperparam-k-metrics}}).

\section{Benchmarking SIE Methods for Viral Mutation Extraction}
\begin{figure}[ht]
    \centering
    \includegraphics[width=\linewidth]{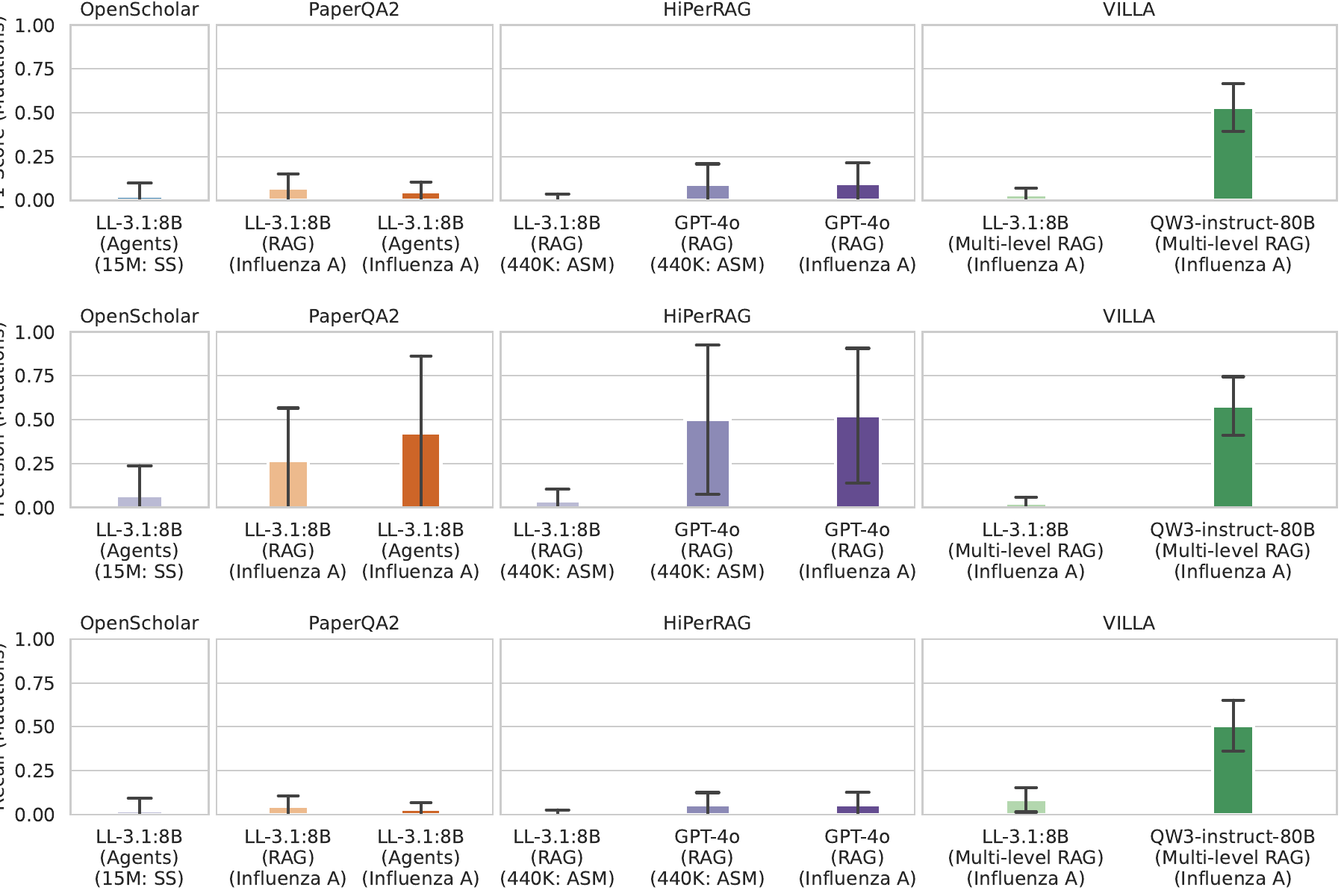}
    \caption[Benchmarking state-of-the art RAG-based tools in viral mutation extraction.]{Distribution of the precision, recall, and F1-scores of OpenScholar, PaperQA2, HiPerRAG, and \ourframework frameworks in retrieving mutations of influenza A viral proteins. The \textit{y}-axis denotes the different configurations of frameworks analysed. Each subplot corresponds to one type of SIE method with different configurations represented using different shades. Bars with lighter shade correspond to flavors of methods with Llama 3.1:8B as the response generator (responder). Bars with darker shades denote the performance of the best configuration for the respective method namely, full agentic framework for paperQA2, GPT-4o as responder for HiPerRAG, and Qwen3-Next-80B-A3B-Instruct as responder for \ourframework. Each bar height and error bar corresponds to the mean and standard deviation of the distribution respectively.}
    \Description{}
    \label{appndx_fig:sota-frameworks-metrics}
\end{figure}

\ourframework performed significantly better than zero-shot prompting, RAG with abstracts, and RAG with full-text when we used nine different LLMs as the response generators.
Our proposed method also outperformed state-of-the-art RAG- and agent-based tools namely, OpenScholar~\citep{asai-hajishirzi-openscholar-arxiv-2024}, PaperQA2~\citep{skarlinski-white-paperqa2-arxiv-2024}, and HiPerRAG~\citep{gokdemir-ramanathan-hiperrag-pascconf-2025}~(\suppfig{\ref{appndx_fig:sota-frameworks-metrics}}).
\begin{figure}[ht]
    \centering
    \includegraphics[width=\linewidth]{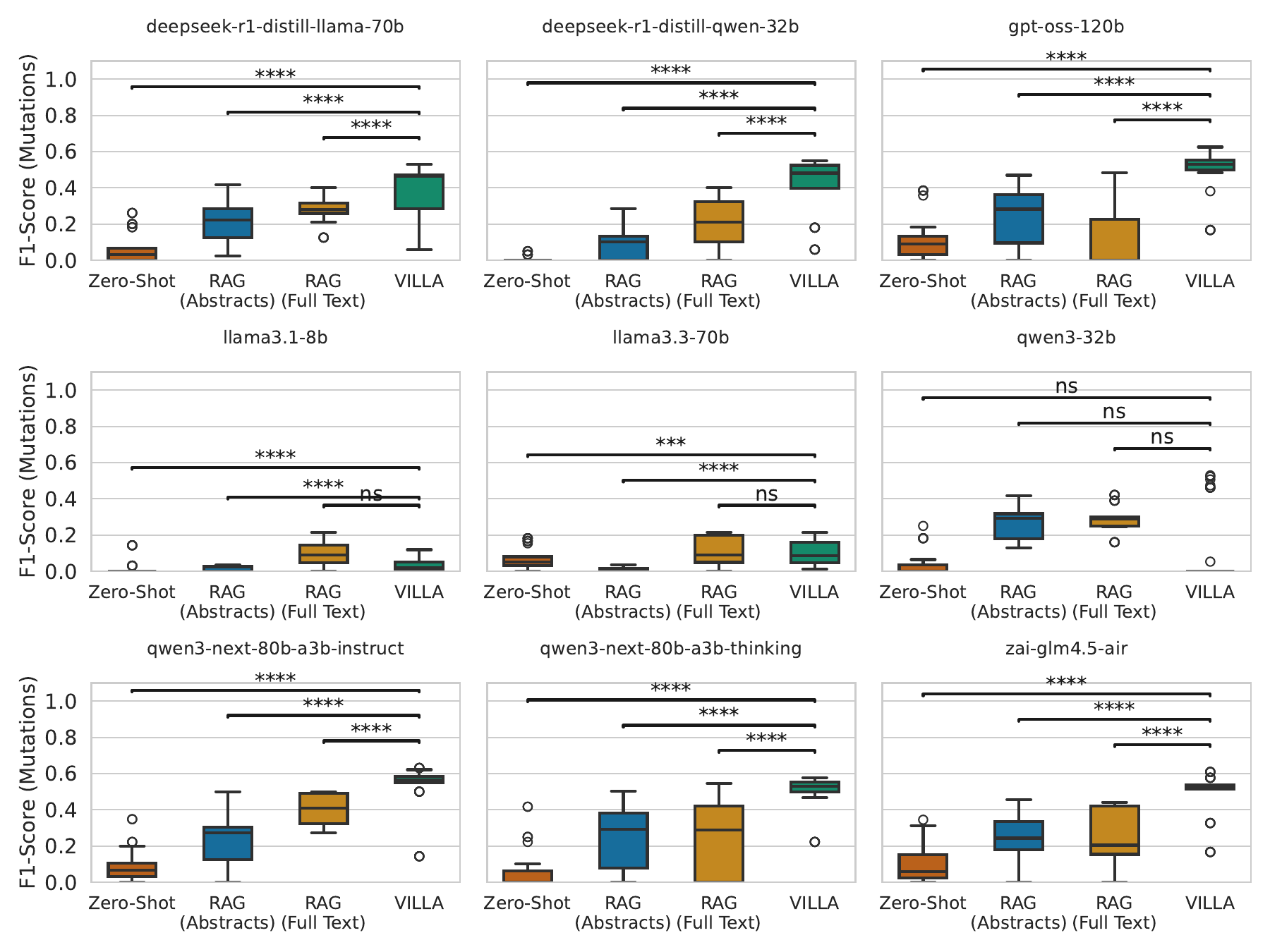}
    \caption{Comparison of F1-scores in SIE of mutations in influenza A virus using zero-shot prompting, RAG with abstracts, RAG with full text, and \mlr methods. Each sub-plot corresponds to a comparison of the methods using a different LLM for response generation. Each box-plot is a distribution of the F1-scores from five iterations of retrieving mutations in ten influenza A proteins using ten different prompts. The stars and `ns' denote the statistical significance computed using Mann-Whitney U test as follows, $p$-value $\le 1\text{e}{-4}$:`****'; $\le 1\text{e}{-4}<$  $p$-value $\le 1\text{e}{-3}$:`***'; $\le 1\text{e}{-3}<$  $p$-value $\le 1\text{e}{-3}$:`**'; $\le 1\text{e}{-2}<$  $p$-value $\le 5\text{e}{-3}$:`*'; $p$-value $> 5\text{e}{-2}$:`ns' (not significant).}
    \Description{}
    \label{appndx_fig:results-comparison-f1-score-all-villa-models}
\end{figure}

\end{document}